\documentclass[journal]{IEEEtran}
\usepackage{makecell}
\usepackage{array}
\usepackage{adjustbox} % 添加 adjustbox 包解决垂直居中问题
\usepackage{tabularx}
\usepackage{graphicx}
\usepackage{xcolor}
\usepackage{multirow}
\usepackage{bm}
\usepackage{amsmath}
\usepackage{booktabs}
\usepackage{array}%newly addled by Kaige
\usepackage{color,soul}
\usepackage{enumerate}
\usepackage{subfigure}
\usepackage{epstopdf} %converting to PDF
\ifCLASSINFOpdf
\usepackage{xcolor}
\input epsf
\usepackage{graphicx}
\usepackage{multicol}
\usepackage{lipsum} % 用于生成占位文本
\usepackage{multirow}
\usepackage{bm}
\usepackage{amsmath}
\usepackage{booktabs}
\usepackage{array}%newly addled by Kaige
\usepackage{color,soul}
\usepackage{enumerate}
\usepackage{subfigure}
\usepackage{soul}
\soulregister\cite7 % 针对\cite命令
\soulregister\ref7 % 针对\ref命令
\soulregister\pageref7 % 针对\pageref命令
\newcommand{\cmmnt}[1]{}
\usepackage{epstopdf} %converting to PDF
\usepackage{makecell}
\renewcommand{\arraystretch}{1.3}

\begin{document}

\title{Advanced Space Mapping Technique Integrating a Shared Coarse Model for Multistate Tuning-Driven Multiphysics Optimization of Tunable Filters}

\author{Haitian~Hu,~\IEEEmembership{Student Member,~IEEE,}
      Wei~Zhang,~\IEEEmembership{Member,~IEEE,}
      Feng~Feng,~\IEEEmembership{Senior Member,~IEEE,}
      Zhiguo~Zhang,
	  and Qi-Jun~Zhang,~\IEEEmembership{Life Fellow,~IEEE}
				\vspace{-6mm}
	\thanks{This work was supported in part by the National Natural Science Foundation
		of China under Grant 62375025 and in part by the Fund of State Key Laboratory of IPOC (BUPT) under Grant IPOC2025ZT07.}
    \thanks{Haitian Hu, Wei Zhang, and Zhiguo~Zhang are with the School of Electronic Engineering, Beijing University of Posts and Telecommunications, Beijing 100876, China (e-mail: haitianhu24@bupt.edu.cn; weizhang13@bupt.edu.cn; zhangzhiguo@bupt.edu.cn).}
    \thanks{Feng Feng is with Tianjin Key Laboratory of Imaging and Sensing Microelectronic Technology, School of Microelectronics, Tianjin University, Tianjin, 300072, China (e-mail: ff@tju.edu.cn).}
    \thanks{Qi-Jun Zhang is with the Department of Electronics, Carleton University, Ottawa, ON K1S5B6, Canada (e-mail: qjz@doe.carleton.ca).}
% <-this % stops a space
}
\maketitle

\begin{abstract}
This article introduces an advanced space mapping (SM) technique that applies a shared electromagnetic (EM)-based coarse model for multistate tuning-driven multiphysics optimization of tunable filters. The SM method combines the computational efficiency of EM single-physics simulations with the precision of multiphysics simulations. The shared coarse model is based on EM single-physics responses corresponding to various nontunable design parameters values. Conversely, the fine model is implemented to delineate the behavior of multiphysics responses concerning both nontunable and tunable design parameter values. {The proposed overall surrogate model comprises multiple subsurrogate models, each consisting of one shared coarse model and two distinct mapping neural networks.} The responses from the shared coarse model in the EM single-physics filed offer a suitable approximation for the fine responses in the multiphysics filed, whereas the mapping neural networks facilitate transition from the EM single-physics field to the multiphysics field. Each subsurrogate model maintains consistent nontunable design parameter values but possesses unique tunable design parameter values. By developing multiple subsurrogate models, optimization can be simultaneously performed for each tuning state. Nontunable design parameter values are constrained by all tuning states, whereas tunable design parameter values are confined to their respective tuning states. This optimization technique simultaneously accounts for all the tuning states to fulfill the necessary multiple tuning state requirements. Multiple EM and multiphysics training samples are generated concurrently to develop the surrogate model. Compared with existing direct multiphysics parameterized modeling techniques, our proposed method achieves superior multiphysics modeling accuracy with fewer training samples and reduced computational costs. The validity of the proposed technique is demonstrated through two tunable microwave filter examples.
\end{abstract}


\IEEEpeerreviewmaketitle

\section{Introduction}
\IEEEPARstart{{R}}{{ECENT}} {advances in microwave component design have increasingly emphasized the significance of multiphysics optimization, in which electromagnetic (EM) performance is tightly coupled with thermal and mechanical effects. For instance, coupled EM-thermal-mechanical simulations have been employed to accurately predict frequency shifts in high-power microwave filters caused by thermal stress \cite{zhang2020ecmtmtt}. In addition, surrogate modeling techniques based on neural-network-enabled space mapping have been developed to capture multiphysics interactions while significantly reducing computational cost \cite{wang2023mlmtt}. Space mapping strategies have also proven effective in the optimization of microwave filters under multiphysics constraints \cite{koziel2021mttmapping}.

Microwave tunable filters play an important role in many fields, such as wireless communication, radar, and satellite communication {\cite{Tunable_Filter1}}{-}{\cite{Tunable_Filter55}}. With the rapid development of wireless communication technology and increasing demands, high-performance and flexible microwave tunable filters are urgently needed. Traditional design methods focus on meeting the requirements of a single tuning state at a fixed frequency. In contrast, microwave tunable filters can adapt their frequency responses across different tuning states, meeting the requirements of multiple tuning states. Therefore, the development of efficient design optimization approaches for microwave tunable filters supporting multiple tuning states is of critical significance.

Fusing electromagnetic (EM) simulations and optimization algorithms is a common approach in the design of microwave tunable filters. However, the direct utilization of EM simulations in design optimization can be computationally demanding, primarily because of the need for numerous repetitive EM simulations to fine-tune design parameters during the optimization process \cite{EM1}-\cite{EM4}. In recent years, artificial neural networks (ANNs) have emerged as powerful tools for parametric modeling and design optimization of microwave filters \cite{ANN3}-\cite{ANN7}. ANNs are proficient at quickly and accurately learning complex input-output relationships. Well-trained ANN models can effectively replace time-consuming direct EM simulations in the design optimization process of microwave filters. 
The combination of space mapping (SM) techniques \cite{SM2}-\cite{SM4} and ANN models has been widely used to accelerate microwave filter optimization. The SM technique assumes the existence of both a coarse model and a fine model \cite{SM7}-\cite{SM9}. The coarse model (e.g., equivalent circuit model) has high computational efficiency but low accuracy, whereas the fine model (e.g., 3D EM simulator) has high accuracy but requires many computational resources. The SM technique exploits the learning ability of ANNs to build mathematical relationships between the coarse and fine models \cite{SM10}, leveraging the accuracy of the fine model with the efficiency of the coarse model. Therefore, the SM technique can efficiently implement computationally intensive EM optimizations via an approximate surrogate model (i.e., coarse model) \cite{SM13}-\cite{SM20}. {Various space mapping-based methods have been developed for multiphysics modeling and optimization of microwave filters. These include EM-centric frameworks combining neural networks to reduce simulation cost \cite{SM1}, parallel data-driven surrogate models with trust-region strategies for high-power designs \cite{SM21}, and comprehensive reviews summarizing recent advances and applications in high-power scenarios \cite{SM0}. A simplified coarse model without tunable elements was also proposed to accelerate tunable filter optimization \cite{SM22}. Efficient modeling under EM-thermal-mechanical coupling has been achieved through the combination of space mapping and neural network surrogates \cite{SM23}. However, these researches only focus on one filter design, which are not applicable to the multistate tuning design (i.e., multiple filters).} {\cite{SM1}}{-}{\cite{SM23}} {can only handle the filter design problem when there is only one design specification. In this article, an advanced space-mapping technique is introduced, which utilizes a shared coarse model and multiple mapping neural networks. Our proposed method focuses on the challenges of microwave tunable filter design to simultaneously optimize multiple sets of tuning parameters to satisfy the specifications for multiple tuning states.}

This article is a significant advance over the work of \cite{Tunable_Filter6} in an effort to further improve the optimization efficiency of the multistate tuning-driven multiphysics optimization of tunable filters. An advanced SM technique that integrates a shared coarse model is proposed for the multistate tuning-driven multiphysics optimization of tunable filters. {The method proposed in this article greatly reduces the multiphysics training sample data by upgrading the subsurrogate model in \cite{Tunable_Filter6} from an ANN to a combination of a shared coarse model and two mapping neural networks. Moreover, all subsurrogate models share a common coarse model, further reducing the single-physics training sample data.} This article introduces the application of the SM technique to the multiphysics modeling problem of microwave tunable filters with multiple tuning states. The EM single-physics responses serve as an approximate solution for the multiphysics responses in the multiphysics filed. However, EM single-physics simulations are significantly faster than are multiphysics simulations. In our proposed technique, we employ the SM technique to establish mappings between the EM filed (single-physics field) and the multiphysics filed. Our proposed overall surrogate model is composed of multiple subsurrogate models. Each subsurrogate model includes one shared coarse model and two distinct mapping neural networks. The shared coarse model is constructed via an ANN model to describe the relationship between the responses of the EM single-physics simulation and the nontunable parameters. The two mapping neural networks, namely, design parameter mapping and frequency parameter mapping, are employed to map the EM filed to the multiphysics filed. The design parameter mapping, constructed via an ANN model, correlates nontunable and tunable parameters in the multiphysics filed and nontunable parameters in the EM filed. The frequency parameter mapping, which is also constructed via an ANN model, establishes the relationship between the tunable parameters and the frequency parameter in the multiphysics filed and the frequency parameter in the EM filed. Each subsurrogate model, which shares identical nontunable parameter values and independently possesses distinct tunable parameter values, represents the behavior of multiphysics responses. We employ parallel computing techniques to generate the shared coarse model samples for the EM single-physics filed and overall surrogate model samples for the multiphysics filed. The optimization process of this overall surrogate model aims to determine the optimal solution for a single set of nontunable design parameters and multiple sets of tunable parameters. Compared with \cite{Tunable_Filter6}, the proposed method in this article reduces the amount of multiphysics data while achieving superior accuracy. Consequently, this approach can speed up the design cycle and enhance design efficiency. Once an accurate comprehensive surrogate model is established, it can offer precise and rapid predictions of multiphysics responses for microwave tunable filters with multiple tuning states. The efficacy of the proposed technique is confirmed through examples involving a tunable evanescent mode cavity filter and a tunable four-pole waveguide filter.

\section{Proposed Multiphysics Optimization Technique for Microwave Tunable Filters with Multiple Tuning States}

\subsection{{Proposed Space-Mapping-Based Surrogate Model Containing a Shared Coarse Model}}
To simultaneously determine an optimal solution set for nontunable design parameters and multiple sets of optimal solutions for t
unable design parameters, ensuring all simultaneously satisfy multiple tuning states for the microwave tunable filter with multiple tuning states, we propose a space-mapping-based multiphysics surrogate model containing a shared coarse model. {All the subsurrogate models 
	share a coarse model, named shared coarse model, which is shared by all the tuning state tasks and mainly used for describing the EM single-physics simulation.} The shared coarse model greatly reduces the number of repeated EM single-physics simulations during the training and optimization stages of the overall surrogate model. The structure of this proposed surrogate model for the microwave tunable filter is illustrated in Fig.~\ref{Overall_Surrogate_Model}. The overall surrogate model comprises multiple subsurrogate models, the total count of which corresponds to the number of required tuning states. Each subsurrogate model consists of one shared coarse model and two distinct mapping neural networks. The responses of the shared coarse model in the EM single-physics filed provide an effective approximation of the fine model responses in the multiphysics filed, whereas the mapping neural networks facilitate the transition from the EM single-physics filed to the multiphysics field. Each constructed subsurrogate model is optimized for a corresponding tuning state. Furthermore, each subsurrogate model shares the same nontunable design parameters while maintaining distinct tunable design parameters. Consequently, the nontunable design parameters are constrained by all the tuning states, whereas the tunable design parameters are constrained by their respective tuning states.

\begin{figure*}[t]
\centering
\includegraphics[width=7.2in]{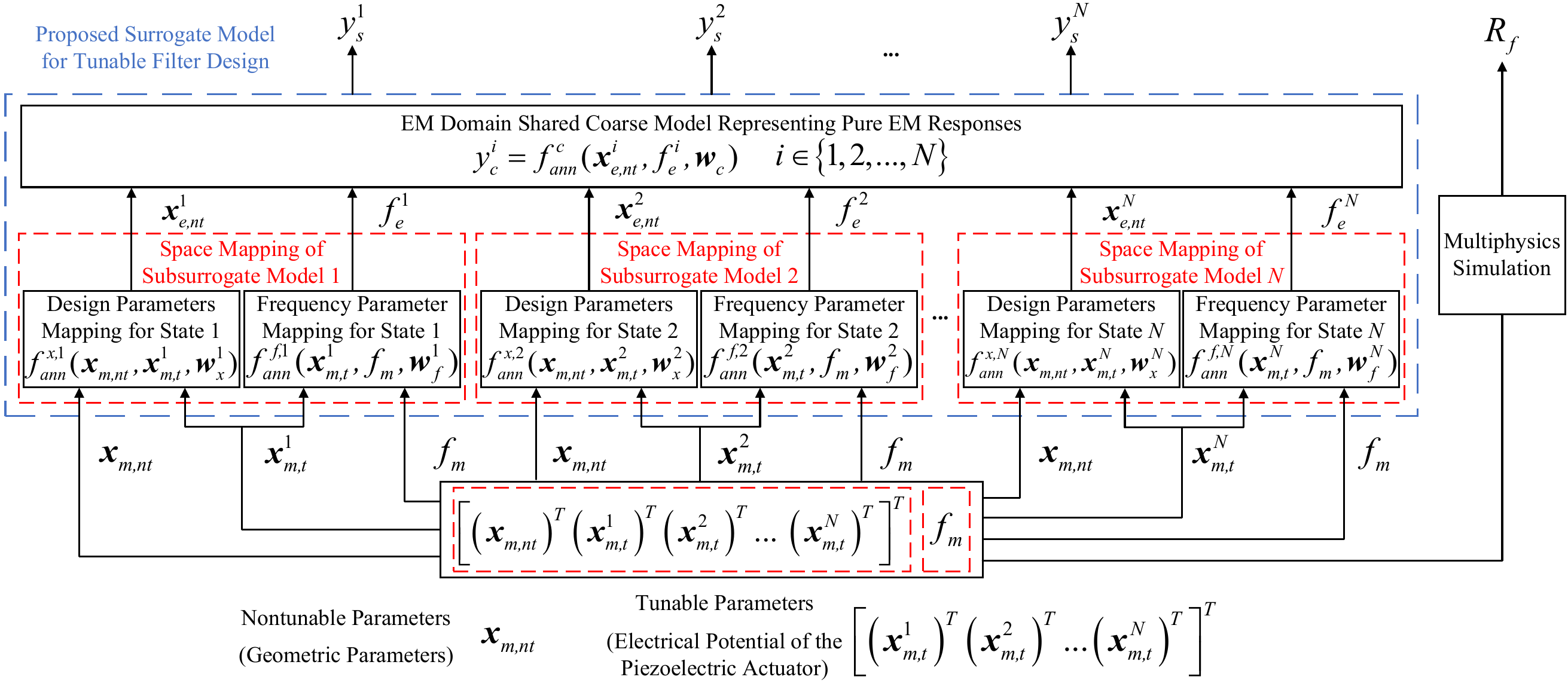}
\caption{Structure of the proposed surrogate model for microwave tunable filter optimization with multiple sets of tuning states.}
\label{Overall_Surrogate_Model}
\end{figure*}

{The shared coarse model represents the relationship between geometric parameters and the output responses of the filter in the EM-only filed.} We denote the $i$th input parameters of the shared coarse model as $\bm{x}^i_{e,nt}$, and the frequency parameter, denoted as $f^i_e$, serves as an additional input parameter. {$i$ denotes the index of the tuning state. $\bm{y}_c$ represents the prediction of all output responses of the tunable filter in the EM-only filed by this shared coarse model. $y^i_c$ represents the prediction of the output response of the tunable filter in the EM-only filed under the $i$th tuning state by the shared coarse model.} $y^i_c$  represented as
\begin{equation}
	y^i_c = f^c_{ann}(\bm{x}^i_{e,nt},f^i_e,\bm{w}_c)
	\label{Coarse_Model}
\end{equation}
where $\bm{w}_c$ denotes a vector that encompasses all the weight parameters of the shared coarse model. $f^c_{ann}$ represents the neural network function. This model delineates the relationship between the responses in the EM single-physics (EM-only) filed and nontunable parameters. 

We propose employing design parameter mapping and frequency parameter mapping to delineate the relationship between EM single-physics (EM-only) filed parameters and multiphysics filed parameters. The objective is to minimize the misalignment between the EM single-physics shared coarse model and the multiphysics fine model. Given the nonlinearity and unpredictability of the relationship between the EM filed parameters and the multiphysics filed parameters, we advocate for the application of an ANN to learn this relationship. We define $f^{x,i}_{ann}$ as the design parameter mapping function represented in the $i$th subsurrogate model. $\bm{x}_{m,nt}$ represents nontunable design parameters in the multiphysics filed, which are consistent across all subsurrogate models. Conversely, $\bm{x}^i_{m,t}$ denotes the tunable design parameters in the multiphysics fileds, specifically for the $i$th subsurrogate model relative to its corresponding tuning state. An ANN model is then trained to learn the relationship between $\bm{x}^i_{e,nt}$ in the EM filed and $\bm{x}_{m,nt}$ and $\bm{x}^i_{m,t}$ in the multiphysics filed, thereby establishing the design parameter mapping. The design parameter mapping for the $i$th subsurrogate model is thus defined as follows:
\begin{equation}
\bm{x}^i_{e,nt} = f^{x,i}_{ann}(\bm{x}_{m,nt},\bm{x}^i_{m,t},\bm{w}^i_x)
\label{Design_Parameter_Mapping}
\end{equation}
where $\bm{x}_{m,nt}$ and $\bm{x}^i_{m,t}$ are denoted as input parameters, while the vector $\bm{w}^i_x$ contains all the weight parameters of the mapping neural network. The output is represented as $\bm{x}^i_{e,nt}$. Similarly, the frequency parameter mapping function for the $i$th surrogate model is denoted as $f^{f,i}_{ann}$. The frequency parameter mapping is constructed through an ANN model captures the relationship between $f^i_e$ from the EM single-physics (EM only) filed and both $\bm{x}^i_{m,t}$ and $f_m$ from the multiphysics field. {Let $f_m$ denote the frequency range used in the simulation process.} The frequency parameter mapping for the $i$th subsurrogate model is defined as
\begin{equation}
f^i_e = f^{f,i}_{ann}(\bm{x}^i_{m,t},f_m,\bm{w}^i_f)
\label{Frequency_Parameter_Mapping}
\end{equation}
where $\bm{x}^i_{m,t}$ represents the inputs. $f_m$ is an additional input that serves as a common parameter for frequency mappings across all subsurrogate models. The output is denoted by $f^i_e$. Furthermore, $\bm{w}^i_f$ denotes a vector containing all the weight parameters of the mapping neural network.

Let $\bm{x}^i_{m,s}$ represent the multiphysics design parameters of the $i$th subsurrogate model that contains the nontunable parameters $\bm{x}_{m,nt}$ and the tunable parameters $\bm{x}^i_{m,t}$, represented as
\begin{equation}
\bm{x}^i_{m,s} = \left[ (\bm{x}_{m,nt})^T\;(\bm{x}^i_{m,t})^T \right]^T
\label{Sub_x}
\end{equation}

{Let $y^i_s$ represent the prediction of the output response of the tunable filter in the multiphysics filed by the ith subsurrogate model}, which is formulated as follows:
\begin{equation}
y^i_s = f^c_{ann}(f^{x,i}_{ann}(\bm{x}_{m,nt},\bm{x}^i_{m,t},\bm{w}^i_x),f^{f,i}_{ann}(\bm{x}^i_{m,t},f_m,\bm{w}^i_f),\bm{w}_c)
\label{Sub_y}
\end{equation}

The design parameters for the overall surrogate model are derived from those of $N$ individual subsurrogate models, represented as
\begin{equation}
\bm{x}_m = \left[ (\bm{x}_{m,nt})^T\;(\bm{x}^1_{m,t})^T\;(\bm{x}^2_{m,t})^T,\cdots,\;(\bm{x}^N_{m,t})^T \right]^T
\label{Overall_x}
\end{equation}

{Let $\bm{y}$ represent the prediction of all output responses of the tunable filter in the multiphysics filed by the overall surrogate model}, which is represented as follows:
\begin{equation}
	\bm{y} \hspace{-1mm}=\hspace{-2mm}
	\left[
	\begin{aligned}
		& f^c_{ann}\hspace{-1mm}\left(f^{x,1}_{ann}(\bm{x}_{m,nt}, \bm{x}^1_{m,t}, \bm{w}^1_x),
		f^{f,1}_{ann}(\bm{x}^1_{m,t}, f_m, \bm{w}^1_f), \bm{w}_c\right) \\[1mm]
		& f^c_{ann}\hspace{-1mm}\left(f^{x,2}_{ann}(\bm{x}_{m,nt}, \bm{x}^2_{m,t}, \bm{w}^2_x),
		f^{f,2}_{ann}(\bm{x}^2_{m,t}, f_m, \bm{w}^2_f), \bm{w}_c\right) \\
		& \;\;\;\;\;\;\;\;\;\;\;\;\;\;\;\;\;\;\;\;\;\;\;\;\;\;\;\;\;\;\;\;\;\;\;\;\;\;\;\;\;\;\; \vdots \\
		& f^c_{ann}\hspace{-1mm}\left(f^{x,N}_{ann}(\bm{x}_{m,nt}, \bm{x}^N_{m,t}, \bm{w}^N_x),
		f^{f,N}_{ann}(\bm{x}^N_{m,t}, f_m, \bm{w}^N_f), \bm{w}_c\right)
	\end{aligned}
	\hspace{-0.5mm}\right]
	\label{Overall_y}
\end{equation}

\subsection{{Parallel Data Generation Process for the EM and Multiphysics Fileds}}
To optimize the microwave tunable filter across multiple tuning states, the space-mapping-based multiphysics surrogate model proposed in Section II.A is employed. We need to develop a surrogate model. The first step in developing this model is to generate training samples. On the basis of the structure of the proposed surrogate model, EM single-physics field shared coarse model data and the multiphysics filed of each subsurrogate model dataset need to be generated. The evaluation of both the EM and multiphysics fileds requires significant computational resources. Moreover, the proposed optimization technique requires multiple iterations to identify optimal solutions that satisfy the required states. The generation of multiphysics training data for each subsurrogate model is completely independent. Consequently, the multiphysics training samples for each subsurrogate model are generated via the COMSOL MULTIPHYSICS 6.1 simulator in parallel across multiple computers with multiple processing cores, significantly reducing time costs. {COMSOL Multiphysics 6.1 is a widely used and well-established commercial Finite Element Analysis software. Its accuracy has been widely verified by benchmarks and comparisons to experimental data in numerous published works (e.g. \cite{COMSOL1},\cite{COMSOL2}). We use a fine mesh setting and perform a mesh convergence analysis to ensure numerical stability and accuracy of the solution. In addition, the physical model and boundary conditions are set strictly according to the theoretical expectations and previous experimental configurations reported in the article. Even if COMSOL Multiphysics 6.1 is not accurate enough, this does not affect our article, because the key point of our proposed method is to reduce the gap between surrogate models and multiphysics simulations. Then, the trained surrogate model can be used instead of the computationally expensive multiphysics simulation to optimize tunable filters with multiple tuning states.} The training samples for the EM single-physics filed shared coarse model are generated via the ANSYS HFSS EM simulator. 

Various distribution methods exist for generating design parameter samples, with the star distribution, grid distribution, and orthogonal distribution being the most commonly employed. In our methodology, we utilize the orthogonal distribution, a specific variant of the design of experiment (DOE) sampling distribution \cite{MP1}, to generate design parameter samples for both the shared coarse model and each subsurrogate model. The orthogonal distribution ensures that subspace divisions are sampled at an identical density, leading to orthogonal subspaces and enhancing the accuracy of the surrogate model. Compared with the star distribution, the orthogonal distribution offers superior accuracy for surrogate models over a broader area. Furthermore, in contrast to the grid distribution, the orthogonal distribution around the central point requires significantly fewer sampling points.

Let $k$ represent the iteration index in the optimization process. {Iteration index is the number of optimization rounds required to achieve convergence conditions during the use of the proposed optimization algorithm}. All the subsurrogate models share a coarse model. Let $\bm{x}^k_{e,nt,c}$ and $\bm{X}^k_e$ denote the central point and training region for the shared coarse model in the EM single-physics filed at the $k$th iteration. Similarly, let $\bm{x}^{i,k}_{m,s,c}$ and $\bm{X}^{i,k}_{m,s}$ represent the central point and training region for the $i$th subsurrogate model in the multiphysics field at the $k$th optimization iteration. Let $n_e$ and $n_m$ represent the total number of EM single-physics samples and multiphysics samples utilized via DOE within one iteration process, respectively. $\bm{X}^k_e$ and $\bm{X}^{i,k}_{m,s}$ are defined via the trust region algorithm as
\begin{equation}
	\bm{X}^k_e = \left\{ \bm{x}_{e,nt} \,\middle|\, \bm{x}^k_{e,nt,c} - \bm{\delta}^k_e \leq \bm{x}_{e,nt} \leq \bm{x}^k_{e,nt,c} + \bm{\delta}^k_e \right\}
	\label{Coarse_X}
\end{equation}
\begin{equation}
	\bm{X}^{i,k}_{m,s} = \left\{ \bm{x}^i_{m,s} \,\middle|\, \bm{x}^{i,k}_{m,s,c} - \bm{\delta}^{i,k}_{m,s} \leq \bm{x}^i_{m,s} \leq \bm{x}^{i,k}_{m,s,c} + \bm{\delta}^{i,k}_{m,s} \right\}
	\label{Sub_X}
\end{equation}
where $\bm{\delta}^k_e$ and $\bm{\delta}^{i,k}_{m,s}$ denote the trust regions of the shared coarse model and the $i$th subsurrogate model, respectively. These parameters delineate the variation range for the design parameters around their respective center points, $\bm{x}^{i,k}_{e,nt,c}$ and $\bm{x}^{i,k}_{m,s,c}$, during the $k$th iteration. {The trust radius defines the local variation range around the current center point for each design variable. The shared coarse model and the $i$th subsurrogate models are trained using samples from the local region centered at $\bm{x}^{i,k}_{e,nt,c}$ and $\bm{x}^{i,k}_{m,s,c}$, respectively. Specifically, the size of the trust region of the coarse model is determined by its corresponding trust radius $\bm{\delta}^k_e$. The size of the trust region of the $i$th subsurrogate model is determined by its corresponding trust radius $\bm{\delta}^{i,k}_{m,s}$.} Once the training regions for all the subsurrogate models are established, the nontunable design parameters obtained from any given subsurrogate model and the combined tunable design parameters from all the subsurrogate models constitute the overall surrogate model training region. This region is computed as follows:
\begin{equation}
	\bm{X}^k_m = \left\{ \bm{x}_m \,\middle|\, \bm{x}^k_{m,c} - \bm{\delta}^k_m \leq \bm{x}_m \leq \bm{x}^k_{m,c} + \bm{\delta}^k_m \right\}
	\label{Overall_X}
\end{equation}
where $\bm{X}^k_m$ represents the training region for the overall surrogate model, $\bm{\delta}^k_m$ denotes the trust region of the overall surrogate model, and $\bm{x}^k_{m,c}$ represents the central point of the overall surrogate model.

To achieve a more precise multiphysics field surrogate model, the training region of the shared coarse model extends beyond that of each subsurrogate model, i.e., $\bm{X}^k_e > \mathop {\max }\limits_{i \in \{ 1,2, \cdots ,N\} } \{ \bm{X}^{i,k}_{m,s}\}$. To increase the efficiency of the proposed multiphysics surrogate optimization, the number $n_e$ of EM single-physics filed shared coarse model samples is selected to be greater than the number $n_m$ of multiphysics filed subsurrogate model samples, i.e., $n_e > n_m$. Fig. \ref{Orthogonal_Distribution} provides a detailed explanation of the orthogonal distribution used to generate multiple design parameter samples around $\bm{x}^k_{e,nt,c}$ and $\bm{x}^{i,k}_{m,s,c}$ for both the shared coarse model and the $i$th subsurrogate model separately at the $k$th optimization iteration. As shown in Fig. \ref{Orthogonal_Distribution}, the shared coarse model and the subsurrogate model share an initial point. The input parameters of the shared coarse model are geometric parameters, and the input parameters of the subsurrogate model are geometric parameters and nongeometric parameters. Once the initial point of the subsurrogate model is determined, we can obtain the values of the geometric parameters as the initial point of the shared coarse model. 

\begin{figure}[!t]
\centering
\includegraphics[width=3.4in]{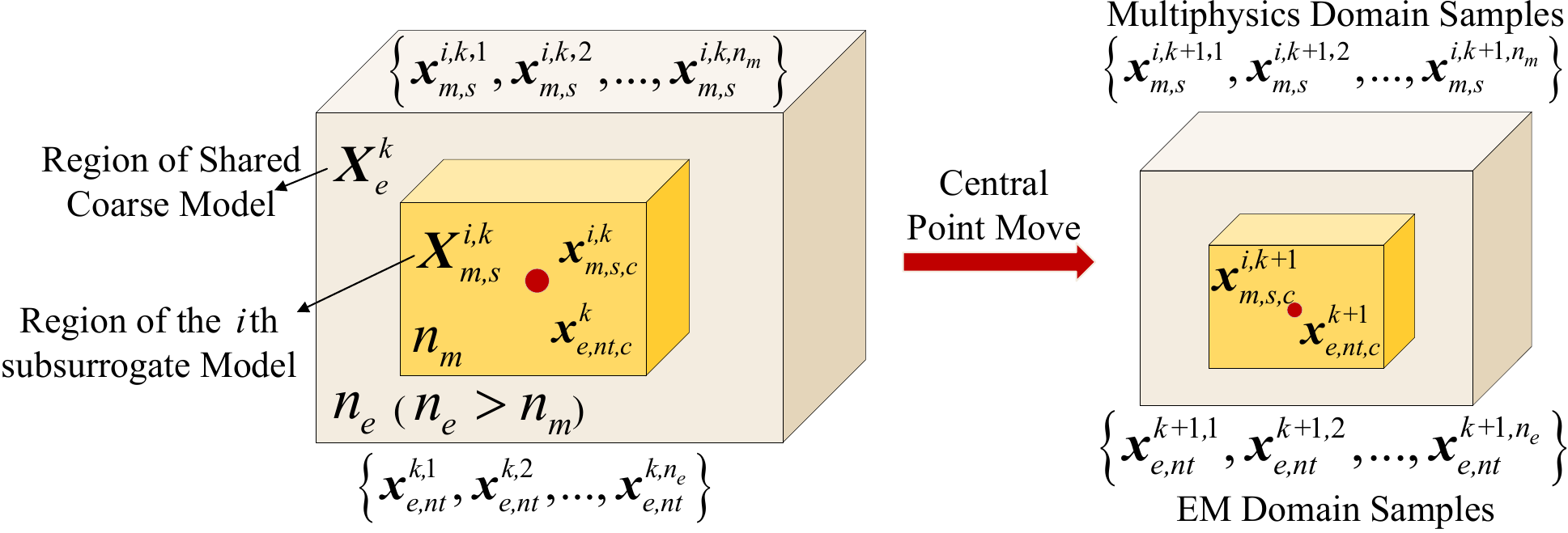}
\caption{Generation of design parameter samples for the EM single-physics (EM-only) shared coarse model and the $i$th multiphysics filed subsurrogate model via the DOE sampling method. Following surrogate optimization at the $k$th iteration, the central points $\bm{x}^k_{e,nt,c}$ and $\bm{x}^{i,k}_{m,s,c}$ move to new updated central points $\bm{x}^{k+1}_{e,nt,c}$ and $\bm{x}^{i,{k+1}}_{m,s,c}$, respectively. Additionally, all orthogonal samples concurrently shift in alignment with these central points.}
\label{Orthogonal_Distribution}
\end{figure}

In our proposed technique, we generate responses for the EM single-physics filed shared coarse model $R_c(\bm{x}^{k,p}_{e,nt},f_e)$ for $p \in \{1, 2, \cdots, n_e\}$ via the ANSYS HFSS EM simulator. $R_c(\bm{x}^{k,p}_{e,nt},f_e)$ represents the EM single-physics response of the shared coarse model. We employ a parallel technique to concurrently generate the responses of all the subsurrogate models in the multiphysics field within the COMSOL MULTIPHYSICS 6.1 simulator. {$R_f(\bm{x}^{k,q}_{m,nt},\bm{x}^{i,k,q}_{m,t},f_m)$ for $q \in \{1, 2, \cdots, n_m\}$ represent the high precision multiphysics filed responses of the tunable filter in COMSOL MULTIPHYSICS 6.1 simulator.} $R_c(\bm{x}^{k,p}_{e,nt},f_e)$ and $R_f(\bm{x}^{k,q}_{m,nt},\bm{x}^{i,k,q}_{m,t},f_m)$ are calculated as
\begin{align}
	\{R_c&(\bm{x}^{k,p}_{e,nt},f_e) \hspace{1.5mm}|\hspace{1.5mm} p = 1,2,\cdots,n_e \} \notag \\
	= \{ &R_c(\bm{x}^{k,1}_{e,nt},f_e), R_c(\bm{x}^{k,2}_{e,nt},f_e), \cdots, R_c(\bm{x}^{k,n_e}_{e,nt},f_e) \}
	\label{Coarse_R}
\end{align}
\begin{align}
	\{R_f&(\bm{x}^{k,q}_{m,nt},\bm{x}^{i,k,q}_{m,t},f_m) \hspace{1.5mm}|\hspace{1.5mm} i = 1,2, \cdots ,N, q = 1,2, \cdots ,n_m \} \notag \\
	= \{&R_f(\bm{x}^{k,1}_{m,nt},\bm{x}^{1,k,1}_{m,t},f_m), R_f(\bm{x}^{k,2}_{m,nt},\bm{x}^{2,k,2}_{m,t},f_m), \notag \\
	&\cdots, R_f(\bm{x}^{k,n_m}_{m,nt},\bm{x}^{N,k,n_m}_{m,t},f_m) \}
	\label{Sub_R}
\end{align}

{The subsurrogate model requires a large amount of multiphysics filed data for training in \cite{Tunable_Filter6}. The proposed method employs a shared coarse model in conjunction with SM. The shared coarse model is trained using  EM single-physics filed data, while the subsurrogate model is trained using multiphysics filed data. Although the EM single-physics filed data also consumes computational resources, it is far less than the resources consumed by multiphysics filed data. Moreover, the responses from the shared coarse model in the EM single-physics filed offer a suitable approximation for the response of the fine model in the multiphysics filed, so we only need a small amount of multiphysics filed data to train the subsurrogate model. Given that all subsurrogate models share a coarse model, this technique significantly reduces the sample size and conserves computer resources. Consequently, this proposed technique accelerates the process of achieving optimal solutions compared with that achieved via the direct utilization of ANN models for multiphysics optimization methods.}

\subsection{{Training Strategy for the Space-Mapping-Based Surrogate Model}}
{After the samples are generated, we can establish the proposed space-mapping-based multiphysics surrogate model. An ANN model is trained with the EM single-physics filed training samples as the shared coarse model. The training process is to optimize the weighted parameters $\bm{w}_c$ of the ANN model in (\ref{Coarse_Model}). The objective is to minimize the sum of squared errors between the EM responses output by the ANSYS HFSS EM simulator and the outputs of the established shared coarse model at $n_f$ frequency points for $n_e$ samples. Here, $n_f$ represents the number of frequency points. This sum of squared errors is denoted as ${E^k_c}(\bm{w}_c)$ and is formulated as}
\begin{equation}
\begin{split}
{E^k_c}(\bm{w}_c) = {\sum\limits_{p = 1}^{n_e} \hspace{-0.5mm}{\sum\limits_{j = 1}^{n_f}\hspace{-0.5mm} {{\left\| {{f^c_{ann}}(\bm{x}^{k,p}_{e,nt},f^j_e,\bm{w}_c) \hspace{-1mm}-\hspace{-1mm} R_c(\bm{x}^{k,p}_{e,nt},f^j_e)} \right\|}^2} } }
\label{Error_Coarse}
\end{split}
\end{equation}
\begin{equation}
\bm{w}^k_c = \arg \mathop {\min }\limits_{\bm{w}_c} \;{E^k_c}(\bm{w}_c)
\label{w_Coarse}
\end{equation}
{where ${E^k_c}(\bm{w}_c)$ represents the error function for the shared coarse model at the $k$th iteration, while $\bm{w}^k_c$ represents the optimal weighting parameters of the neural network. Once the error function value of the shared coarse model fall below a user-defined threshold (e.g., 10\%), the shared coarse model training terminates. The weight parameters of the trained shared coarse model are fixed in the following process of the $k$th iteration. The developed shared coarse model can offer accurate approximations the following multiphysics surrogate model.}

{The unit mappings for design parameter mappings and frequency parameter mappings are performed by setting the input parameter values of the EM filed equal to those of the multiphysics filed. The purpose of the unit mapping is to provide better initial values for training the mapping neural networks. After the unit mappings are established, the training process of the multiphysics subsurrogate models is carried out to obtain the final overall surrogate model. All subsurrogate models are trained simultaneously using the multiphysics input parameters as model inputs and the multiphysics filed responses as model outputs. During this process, the weight parameters $\bm{w}^i_x$ and $\bm{w}^i_f$ of the mapping modules are optimized to minimize the error between the proposed multiphysics model and the multiphysics training data. Let $\bm{w}^i_s$ include all the weight parameters in the mapping neural networks of the $i$th subsurrogate model, and let ${E^{i,k}_m}(\bm{w}^i_s)$ represent the error function of the $i$th subsurrogate model at the $k$th iteration, which is formulated as}
\begin{equation}
	\begin{aligned}
		E^{i,k}_m(\bm{w}^i_s) = \sum\limits_{q=1}^{n_m} \sum\limits_{j=1}^{n_f}
		\Bigg\| f^c_{ann}\Big(f^{x,i}_{ann}(\bm{x}^{k,q}_{m,nt},\bm{x}^{i,k,q}_{m,t},\bm{w}^i_x), \\
		f^{f,i}_{ann}(\bm{x}^{i,k,q}_{m,t},f^j_m,\bm{w}^i_f), \bm{w}_c \Big)
		- R_f(\bm{x}^{k,q}_{m,nt},\bm{x}^{i,k,q}_{m,t},f^j_m) \Bigg\|^2
		\label{Error_Sub}
	\end{aligned}
\end{equation}
where
\begin{equation}
\bm{w}^i_s = [({\bm{w}^i_x})^T\;({\bm{w}^i_f})^T]^T
\end{equation}

{Let ${E^k_m}(\bm{w})$ represent the error function of the overall multiphysics filed surrogate model at the $k$th iteration, which is formulated as}
\begin{equation}
	\begin{aligned}
		E^k_m(\bm{w}) &= E^{1,k}_m(\bm{w}^1_s) + E^{2,k}_m(\bm{w}^2_s) + \cdots + E^{N,k}_m(\bm{w}^N_s) \\
		&= \sum\limits_{i=1}^N \sum\limits_{q=1}^{n_m} \sum\limits_{j=1}^{n_f}
		\Bigg\| f^c_{ann}\Big(f^{x,i}_{ann}(\bm{x}^{k,q}_{m,nt},\bm{x}^{i,k,q}_{m,t},\bm{w}^i_x), \\
		&\quad f^{f,i}_{ann}(\bm{x}^{i,k,q}_{m,t},f^j_m,\bm{w}^i_f), \bm{w}_c \Big)
		\hspace{-1mm}-\hspace{-1mm} R_f(\bm{x}^{k,q}_{m,nt},\bm{x}^{i,k,q}_{m,t},f^j_m) \Bigg\|^2
		\label{Error_Overall}
	\end{aligned}
\end{equation}
where
\begin{equation}
\bm{w} = [(({\bm{w}^1_s})^T\;({\bm{w}^2_s})^T, \cdots ,\;({\bm{w}^N_s})^T]^T
\label{w_Overall}
\end{equation}
\begin{equation}
\bm{w}^k = \arg \mathop {\min }\limits_{\bm{w}} \;{E^k_m}(\bm{w})
\label{w_Overall_fin}
\end{equation}
{where $\bm{w}^k$ represents all optimal weight parameters within the mapping neural networks of the overall surrogate model. If ${E^k_m}(\bm{w})$ falls below a user-specified threshold (for instance, 2\%), the training process for the overall surrogate model is concluded at the $k$th iteration. The resulting trained surrogate model can then be utilized to optimize the microwave tunable filter across multiple tuning states.}

\subsection{{Optimization Process for the Proposed Surrogate Model with Multiple Tuning States}}
Let $\bm{x}^0_m$ represent the initial point for multiphysics optimization of a microwave tunable filter with multiple tuning states. In the initial iteration of the optimization process, we initialize the iteration counter $k=1$. We set the center point $\bm{x}^k_{m,c} = \bm{x}^0_m$ of the proposed overall surrogate model in the first iteration. We can obtain the central point $\bm{x}^{i,k}_{m,s,c}$ ($i\,\in\,\{1,\,2,\,\cdots,\,N\}$) of the $i$th subsurrogate model from $\bm{x}^k_{m,c}$, which includes nontunable design parameters and the $i$th set of tunable design parameters. The center point $\bm{x}^k_{e,nt,c}$ of the shared coarse model is equal to the center point of the nontunable design parameters of the subsurrogate model.

The training samples for the shared coarse model are generated from EM filed (single-physics filed) simulations using the ANSYS HFSS EM simulator. In each optimization iteration, a parallel computer technique is used to generate all the multiphysics training samples for the subsurrogate models. The overall surrogate model training samples (i.e., multiphysics filed samples of all the subsurrogate models) are derived from the multiphysics filed simulation via the COMSOL MULTIPHYSICS 6.1 simulator. After sample generation, we develop the proposed multiphysics surrogate model for optimizing the microwave tunable filter with multiple tuning states in NeuroModelerPlus \cite{NeuroModeler} software according to the structure shown in Fig.~\ref{Overall_Surrogate_Model}.
{The proposed method is not restricted to this software. All the core procedures described in this article, such as multiple subsurrogate model construction, trust filed optimization, EM evaluation, and so on, are easily implemented in other environments (such as Python environment, e.g., PyTorch or TensorFlow), MATLAB, or other optimization toolboxes, as long as they support surrogate model modeling and EM solver integration. The use of NeuroModelerPlus in this article mainly serves to accelerate the model development and ensure seamless integration between surrogate modeling and EM simulation, as well as between surrogate modeling and multiphysics simulation.} The training process comprises two stage. The first stage involves developing a shared coarse model to represent the relationship between EM single-physics responses and nontunable design parameters, which is defined in Section III.C. Once the shared coarse model is trained, the weight parameters within the shared coarse model are fixed in the following process of the $k$th optimization iteration. The trained shared coarse model provide prior knowledge for the subsequent overall surrogate model. The second stage involves establishing an overall surrogate model using multiphysics filed samples. $N$ subsurrogate models are established to represent the relationship between the multiphysics responses corresponding to different tuning states and the design parameters (i.e., $\bm{x}^i_s$). To expedite the overall surrogate model training process, parallel training methods are employed to train all the subsurrogate models simultaneously. The training of the overall surrogate model is not terminated until all the errors $E^{i,k}_m(\bm{w}^i_s)$ ($i = 1,2, \cdots ,N$) of all the subsurrogate models are less than the user-defined training error threshold. Then, we employ the trained overall surrogate model to optimize the multiphysics design of the microwave tunable filter with multiple tuning states.

We perform multiphysics optimization with $N$ tuning states to find the optimal solution for the next iteration via the proposed surrogate model, which is formulated as
%\vspace{-7mm}
\begin{equation}
{\bm{x}^k_m} = \arg \;\mathop {\hspace{-0mm}\min }\limits_{\bm{x}_m \in \bm{X}^k_m} \hspace{-0mm}U\left( {\sum\limits_{i = 1}^N {{\left\| {y^i_s(\bm{x}_{m,nt},\bm{x}^i_{m,t},f_m) - {S^i}} \right\|}^2} } \right)
\label{overall_xk}
\end{equation}
where $\bm{x}^k_m$ denotes the optimal solution of the overall surrogate model after the $k$th optimization iteration, $\bm{X}^k_m$ represents the training region of the proposed overall surrogate model, ${y^i_s}$ denotes the response of the $i$th subsurrogate model, where $S^i$ denotes the $i$th tuning state. In the optimization process of the overall surrogate model, the nontunable design parameters $\bm{x}_{m,{nt}}$ are constrained by all the subsurrogate models in all the tuning states, whereas the $i$th tunable design parameters $\bm{x}^i_{m,t}$ are constrained only by the $i$th set of tuning state corresponding to the $i$th subsurrogate model.

In this article, we introduce a trust-region algorithm that is specifically tailored to the space-mapping-based multiphysics optimization of a microwave tunable filter with multiple states. The algorithm can enhance the convergence of the proposed multiphysics optimization. Let ${U}({R^i_f}({\bm{x}^k_{m,{nt}}},{\bm{x}^{i,k}_{m,t}}))$ and ${U}({R^i_f}({\bm{x}^{k-1}_{m,{nt}}},{\bm{x}^{i,{k-1}}_{m,t}}))$ represent the objective functions of the multiphysics analysis at ${\bm{x}^{i,k}_s}$ and ${\bm{x}^{i,{k-1}}_s}$, respectively. Let ${U}({y^i_s}({\bm{x}^k_{m,{nt}}},{\bm{x}^{i,k}_{m,t}}))$ and ${U}({y^i_s}({\bm{x}^{k-1}_{m,{nt}}},{\bm{x}^{i,{k-1}}_{m,t}}))$ represent the objective functions of the $i$th subsurrogate model at ${\bm{x}^{i,k}_s}$ and ${\bm{x}^{i,{k-1}}_s}$, respectively. Finally, let $r^{i,k}_{m,s}$ be the control parameter, which represents the ratio of the reduction in the actual objective function and the reduction in the predicted objective function, which is formulated as
\begin{equation}
r^{i,k}_{m,s}\hspace{-0.5mm}=\hspace{-1.5mm}\left\{ \hspace{-1.5mm}
\begin{array}{c}
-1, \hspace{4mm} \text{if} \; {U}({R^i_f}({\bm{x}^k_{m,{nt}}},{\bm{x}^{i,k}_{m,t}})\hspace{-0.5mm}) \hspace{-1mm}> \hspace{-1mm}{U}({R^i_f}({\bm{x}^{k-1}_{m,{nt}}},{\bm{x}^{i,{k-1}}_{m,t}})\hspace{-0.5mm}) \hspace{-1.5mm}  \vspace{2mm}\\
\hspace{-0.5mm}\displaystyle{\frac{{U}({R^i_f}({\bm{x}^k_{m,{nt}}},{\bm{x}^{i,k}_{m,t}})\hspace{-0.5mm}) \hspace{-1mm}-\hspace{-1mm}{U}({R^i_f}({\bm{x}^{k-1}_{m,{nt}}},{\bm{x}^{i,{k-1}}_{m,t}})\hspace{-0.5mm}) \hspace{-1.5mm}}{{U}({y^i_s}({\bm{x}^k_{m,{nt}}},{\bm{x}^{i,k}_{m,t}}\hspace{-0.5mm})\hspace{-0.5mm}) \hspace{-1mm}-\hspace{-1mm}{U}({y^i_s}({\bm{x}^{k-1}_{m,{nt}}},{\bm{x}^{i,{k-1}}_{m,t}})\hspace{-0.5mm}) \hspace{-1.5mm}}},  \text{otherwise}
\end{array} \right.
\label{r}
\end{equation}

The trust radius $\bm{\delta}^{i,k}_{m,s}$ is modified according to $r^i_{m,s}$. $\bm{\delta}^{i,k}_{m,s}$ contains the trust radius $\bm{\delta}^{i,k}_{m,s,nt}$ of the nontunable design parameters and the trust radius $\bm{\delta}^{i,k}_{m,s,t}$ of the tunable design parameters corresponding to the $i$th subsurrogate model. At the $(k+1)$th optimization iteration, the trust radius $\bm{\delta}^{i,{k+1}}_{m,s}$ is derived as
\begin{equation}
\begin{split}
\bm{\delta}^{i,{k+1}}_{m,s}\hspace{-0.5mm}  =\hspace{-0.5mm}  \left\{ {\begin{aligned}
&{\eta_e \bm{\delta}^{i,k}_{m,s},}\hspace{-1mm} &&{r^{i,k}_{m,s}>0.7}\\
&{\eta_c \bm{\delta}^{i,k}_{m,s},}\hspace{-1mm}&&{r^{i,k}_{m,s}<0.4} \\
&{\bm{\delta}^{i,k}_{m,s},}\hspace{-1mm}&&{\text{otherwise}}
\end{aligned}} \right.
\end{split}
\label{delta_sub_1}
\end{equation}
where
\begin{equation}
\bm{\delta}^{i,{k+1}}_{m,s} = \left[ ({\bm{\delta}^{i,{k+1}}_{m,s,{nt}}})^T\;({\bm{\delta}^{i,{k+1}}_{m,s,t}})^T\right]^T
\label{delta_sub_2}
\end{equation}
where $\eta_e$ and $\eta_c$ represent the artificially defined expansion coefficient and contraction coefficient, respectively. {The trust radius is adaptively updated at each iteration based on  (\ref{r})-(\ref{delta_sub_2})}.  Considering the accuracy of the model and optimizing the iteration speed comprehensively, we set $\eta_e=0.7$ and $\eta_c=0.4$. The trust radius of the overall surrogate model is composed of the minimum trust radius corresponding to the nontunable design parameters and the trust radius corresponding to the tunable design parameters of all the subsurrogate models, which is calculated as
\begin{equation}
\bm{\delta}^{k+1}_m\hspace{-0.5mm}  =\hspace{-1mm} \left[ {{{(\bm{\delta}^{k+1}_{m,s,nt})}^T}}{{{(\bm{\delta}^{1,{k+1}}_{m,s,t})}^T}}{{{(\bm{\delta}^{2,{k+1}}_{m,s,t})}^T}} \cdots {{{(\bm{\delta}^{N,{k+1}}_{m,s,t})}^T}}\right]^T
\label{delta_overall}
\end{equation}
where
\begin{equation}
\bm{\delta}^{k+1}_{m,s,nt}= \mathop {\min }\limits_{i \in \{ 1,2, \cdots ,N\} } \{ \bm{\delta}^{i,{k+1}}_{m,s,nt}\}
\label{delta_overall_nt}
\end{equation}

To accurately map multiphysics parameters and EM single-physics parameters, the trust radius of each subsurrogate model in the multiphysics field is set slightly smaller than that of the shared coarse model in the EM single-physics field in (\ref{delta_coarse}). We define $\gamma$ as the ratio between the maximum trust radius among all the subsurrogate models and the trust radius of the shared coarse model. This ratio is generally greater than one. Establishing this relationship strengthens the space-mapping-based surrogate model for multiphysics optimization, resulting in improved robustness and enhanced effectiveness of the proposed multiphysics optimization technique. Once the multiphysics trust radius in the overall surrogate model is established, we can determine the EM single-physics trust radius $\bm{\delta}^{k+1}_e$ of the shared coarse model on the basis of $\gamma$, which is defined as
\begin{equation}
\bm{\delta}^{k+1}_e = \gamma\mathop {\max }\limits_{i \in \{ 1,2, \cdots ,N\} } \{ \bm{\delta}^{i,{k+1}}_{m,s}\}
\label{delta_coarse}
\end{equation}

The multiphysics optimization process for the proposed surrogate model will terminate if either the normalized absolute difference between the design parameters in two consecutive iterations is sufficiently small or if the absolute difference between the sums of the multiphysics objective function values of all the subsurrogate models in two subsequent iterations is sufficiently small. The termination criterion is defined by the following formula:
\begin{equation}
\label{stop1}
\frac{\left\|\bm{x}^k_m - \bm{x}^{k-1}_m \right\|} {\left\|\bm{x}^k_m\right\|} \le \epsilon
\end{equation}
\begin{equation}
\label{stop2}
\sum\limits_{i = 1}^N \left\|{ {U}({y^i_s}({\bm{x}^k_{m,{nt}}},{\bm{x}^{i,k}_{m,t}})) - {U}({y^i_s}({\bm{x}^{k-1}_{m,{nt}}},{\bm{x}^{i,{k-1}}_{m,t}}))} \right\| \le \epsilon
\end{equation}
where $\epsilon$ represents a user-defined threshold (e.g., $10^{-4}$).

Once the termination condition (\ref{stop1}) or (\ref{stop2}) is met, the optimization terminates at the $k$th optimization iteration. We obtain the final optimal multiphysics solution $\bm{x}^k_m$ of the proposed surrogate model for the microwave tunable filter with multiple tuning states. If neither condition is met, the solution from the current iteration is set as starting point for the next optimization iteration, i.e., $\bm{x}^{k+1}_{m,c} = \bm{x}^{k}_m$. We increase the index of the iteration by 1 and start the next optimization iteration. In each optimization iteration, we use low-cost EM single-physics samples to develop shared coarse model, providing valuable prior knowledge for establishing the overall surrogate model. Thus, the proposed technique enables the construction of an effective surrogate model with large neighborhoods. Each multiphysics optimization iteration achieves large-scale and efficient optimization updates. Therefore, this method can be applied to obtain the optimal multiphysics solution with fewer iterations. The flowchart of the proposed tunable ﬁlter optimization technique using a novel surrogate model in this article is shown in Fig.~\ref{Overall_flowchart}. The detailed steps of the optimization algorithm utilizing the proposed surrogate model are outlined below.

\begin{figure*}
\centering
\includegraphics[width=7.2in]{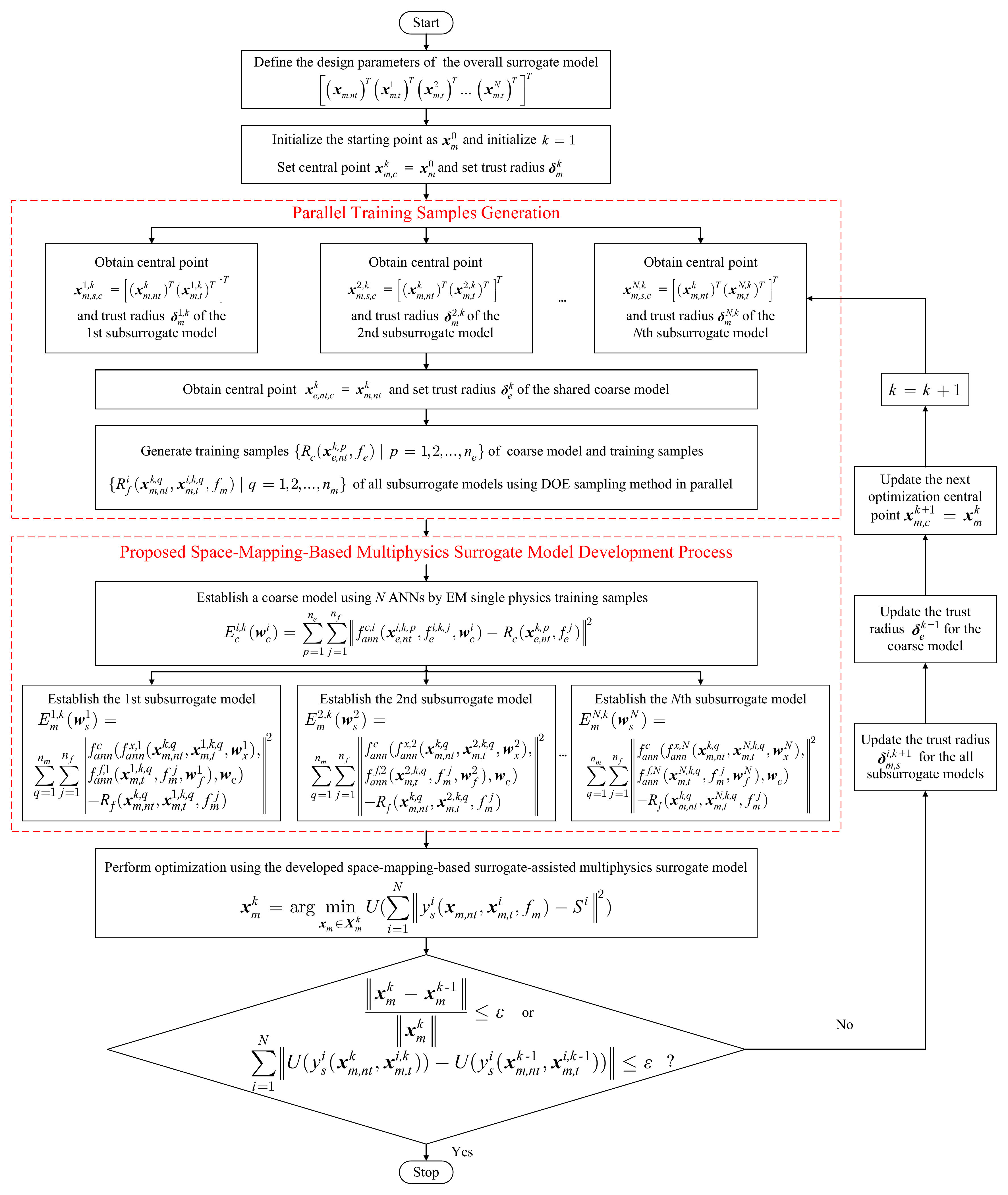}
\caption{Flowchart of the proposed tunable ﬁlter optimization technique using a novel surrogate model.}
\label{Overall_flowchart}
\end{figure*}

\begin{enumerate}[ Step 1)]
\item Define the design parameters of the overall surrogate model for the multiphysics field, i.e., $\bm{x}_m = {\left[ {{(\bm{x}_{m,nt})}^T{(\bm{x}^1_{m,t})}^T{(\bm{x}^2_{m,t})}^T \cdots {(\bm{x}^N_{m,t})}^T} \right]^T}$. Obtain the design parameters $\bm{x}^i_{m,s}= {\left[ {{(\bm{x}_{m,nt})}^T{(\bm{x}^i_{m,t})}^T} \right]^T}$ ($i\,\in\,\{1,\,2,\,\cdots,\,N\}$) for the $i$th subsurrogate model corresponding to the $i$th tuning state from $\bm{x}_m$.
\item Set the starting point as $\bm{x}^0_m$ and initialize $k=1$. Set the initial center point of the overall multiphysics field surrogate model $\bm{x}^k_{m,c}=\bm{x}^0_m$. Initialize the trust radius of this model to $\bm{\delta}^{k}_m$.
\item Obtain the center point of the $i$th subsurrogate model $\bm{x}^{i,k}_{m,s,c}= {\left[ {{(\bm{x}^k_{m,nt})}^T{(\bm{x}^{i,k}_{m,t})}^T} \right]^T}$ from $\bm{x}^k_{m,c}$. Obtain the trust radius $\bm{\delta}^{i,k}_m$ of the $i$th subsurrogate model from $\bm{\delta}^{k}_m$.
\item Obtain the EM single-physics center point for the shared coarse model from $\bm{x}^k_{m,c}$, i.e., $\bm{x}^k_{e,nt,c} = \bm{x}^k_{m,nt}$. Set the EM single-physics trust radius $\bm{\delta}^k_e$ for the shared coarse model using (\ref{delta_coarse}).
\item Generate training samples for the EM single-physics field shared coarse model around the center point $\bm{x}^k_{e,nt,c}$ through (\ref{Coarse_R}) and training samples for all the multiphysics field subsurrogate models around the center point $\bm{x}^{i,k}_{m,s,c}$ through (\ref{Sub_R}) via the DOE sampling method with a parallel technique.
\item Establish a shared coarse model using $n_e$ sets of training samples by optimizing the weighting parameters $\bm{w}^{i,k}_c$ within the ANN model in (\ref{Coarse_Model}) via the parallel technique simultaneously.
\item Establish the overall surrogate model consisting of $N$ subsurrogate models by developing $N$ design parameter mappings and $N$ frequency parameter mappings by optimizing the weighting parameters $\bm{w}$ within the mapping neural networks in (\ref{w_Overall}) via the parallel technique simultaneously.
\item Use the developed surrogate model to perform design optimization and find the next optimal surrogate solution $\bm{x}^{k}_m$ via (\ref{overall_xk}).
\item If either termination conditions in (\ref{stop1}) or (\ref{stop2}) is satisfied, then go to Step (15); otherwise, go to Step (10).
\item Update the trust radius for the $i${th} subsurrogate model $\bm{\delta}^{i,{k+1}}_{m,s}$ via (\ref{r})--(\ref{delta_sub_2}).
\item Update the trust radius for the overall surrogate model $\bm{\delta}^{k+1}_m$ via (\ref{delta_overall})--(\ref{delta_overall_nt}).
\item Update the trust radius for the shared coarse model $\bm{\delta}^{k+1}_e$ via (\ref{delta_coarse}).
\item Update the next optimization center point $\bm{x}^{k+1}_{m,c} = \bm{x}^{k}_m$.
\item Increase the iteration counter $k=k+1$ and go to Step (3).
\item Stop the microwave tunable filter optimization process.
\end{enumerate}

\section{Numerical Examples}
\subsection{Multiphysics Optimization of a Tunable Evanescent-Mode Cavity Filter Using a Piezoactuator}
The first example involves the optimization of a tunable evanescent-mode cavity filter. The structure of this filter, which incorporates a piezoactuator, is depicted in Fig.~\ref{Ex1_Structure} \cite{Ex1}. The piezoactuator is made of lead zirconate titanate (PZT-5H). When an electric field is applied to the piezoactuator, it generates a geometric strain directly proportional to the electric field due to the piezoelectric effect \cite{PE}. The length, width, and height of the filter cavity are fixed at 100 mm, 50 mm, and 50 mm, respectively. $W$ denotes the column width, $L$ represents the column length, and $H$ signifies the gap between the piezoactuator bottom and the column top. The nontunable design parameters for the overall surrogate model are $W$, $L$, and $H$. $V$ denotes the electronic potential applied to the piezoactuator, a multiphysics design parameter. The applied voltage ($V$) will cause deformation and displacement of the piezoactuator, subsequently changing the air gap ($H$), thereby providing tunability for the tunable evanescent mode cavity filter. Fig. \ref{Ex1_Structure_two_point} shows the deformation structures and responses of the filter with the same geometrical size $[{W}\;{L}\;{H}]^T = [16.5178\; 15.2459\; 243.874]^T$ [mm mm um] under varying input voltages. In Fig. \ref{Ex1_Structure_two_point} (a), the piezoactuator deflects downward at an applied voltage of ${V} = 400$ [V]. In Fig. \ref{Ex1_Structure_two_point} (b), the piezoactuator deflects upward at a voltage of ${V} = -400$ [V]. The tunable design parameter for the overall surrogate model is $V$, which can be adjusted within a reasonable range, specifically between -600 V and 600 V. For this filter example, there are two tuning states:
\begin{equation}
\begin{array}{l}
 \text{State 1:}\left| {{S_{11}}} \right| \le -10 \;\text{dB}, \;3.038 \;\text{GHz} \le {{f}} \le 3.042 \;\text{GHz} \vspace{1mm}\\
 \text{State 2:}\left| {{S_{11}}} \right| \le -10 \;\text{dB}, \;3.090 \;\text{GHz} \le {{f}} \le 3.094 \;\text{GHz}
\end{array}
\label{Ex1_tuning_states}
\end{equation}

\begin{figure}
\centering
\includegraphics[width=3in]{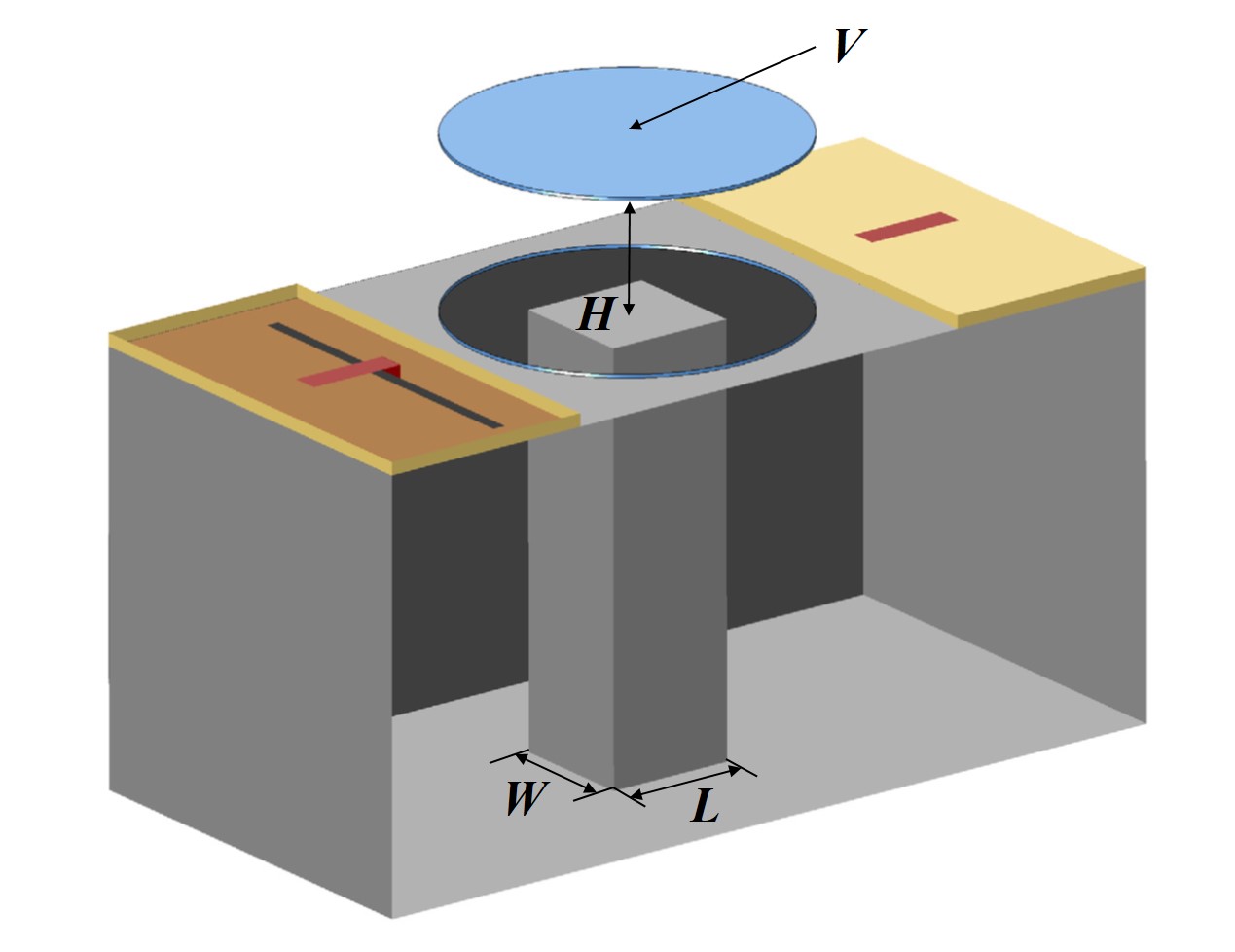}
\caption{Configuration of the tunable evanescent mode cavity filter using a piezoactuator.} \vspace{-3.4mm}
\label{Ex1_Structure}
\end{figure}

\begin{figure}[!t]
\centering
\subfigure[]{\includegraphics[width=3.5in]{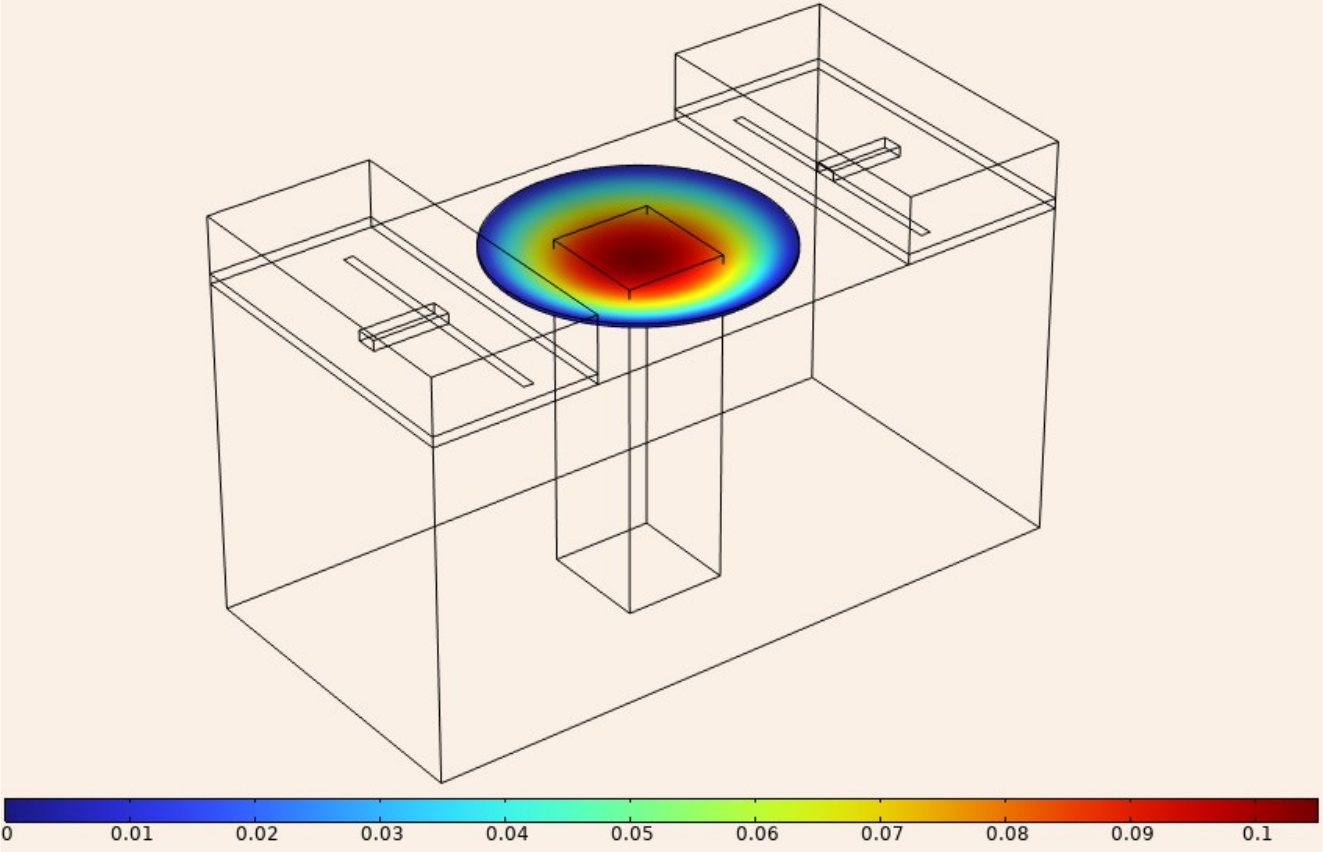}}
\subfigure[]{\includegraphics[width=3.5in]{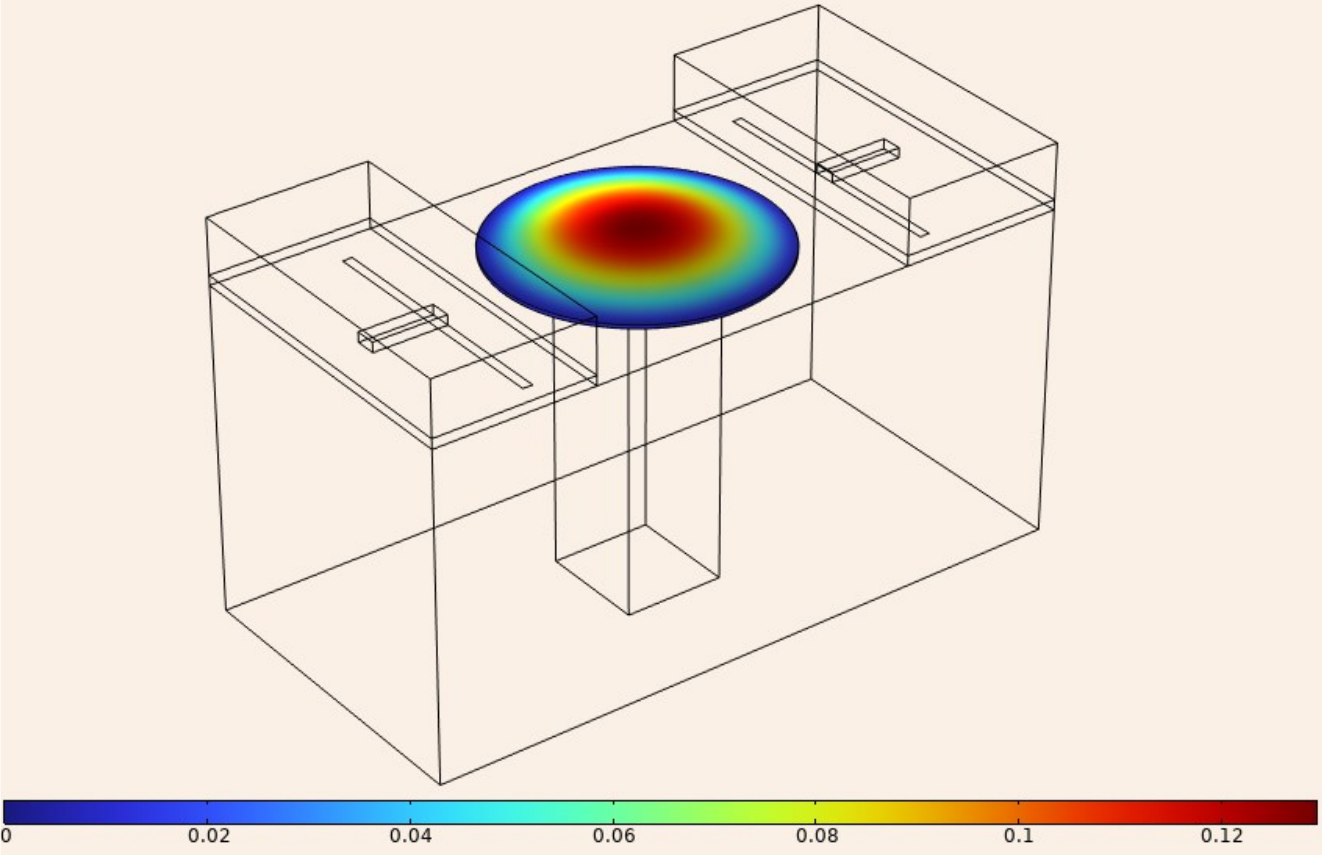}}
\caption{Deformation structure of the tunable evanescent mode cavity filter due to piezoelectric effects. (a) The piezoactuator deflects downward a voltage of ${V} = 400$ [V]. (b) The piezoactuator deflects upward from the bottom with a voltage of ${V} = -400$ [V].} \vspace{-5mm}
\label{Ex1_Structure_two_point}
\end{figure}
As shown in Fig. \ref{V_Po400_Ne400}, the response of the tunable evanescent-mode cavity filter primarily shifts with frequency when the input voltage is changed. Consequently, we propose a overall surrogate model that incorporates two subsurrogate models, as depicted in Fig. \ref{Ex1_Overall_Surrogate_Model}. Both subsurrogate models are constructed from one shared coarse model and two distinct mapping neural networks. Each subsurrogate model shares identical values of nontunable design parameters while independently maintaining different values of tunable design parameters. This approach ensures that each subsurrogate model can accommodate its corresponding tuning state. The input parameters for the shared coarse model are $\bm{x}^1_{e,nt}=[{W}\;{L}\;{H}]^T$ and $\bm{x}^2_{e,nt}=[{W}\;{L}\;{H}]^T$. The frequencies $f^1_e$ and $f^2_e$ serve as additional input parameters for the shared coarse model. The shared coarse model delineates the relationship between the EM single-physics responses $S_{11}$ and the nontunable parameters, thereby providing preliminary knowledge for the overall multiphysics field model. 

{The purpose of this article is multiple filters design simultaneously, each subsurrogate model represents one filter design. Therefore, the values of tuning parameters are different for different subsurrogate models (i.e., different filters). We design two filters simultaneously. There is one voltage $V$ to be the tunable parameter for each filter. $V_1$ and $V_2$ are the controlling voltages for the first filters and second filter, respectively.} The design parameters for the overall surrogate model are defined as
\begin{equation}
\bm{x}_m = [{W}\;{L}\;{H}\;{V_1}\;{V_2}]^T
\end{equation}
{Where $\bm{x}_{m,nt}=[{W}\;{L}\;{H}]^T$ denotes the nontunable parameters with unit [mm mm um]. These parameters serve as the shared input parameters of the both subsurrogate models, which are constrained by the two tuning states simultaneously. The variable $x^1_{m,t}=V_1$ denotes the input parameter of the first subsurrogate model with unit V, and the variable $x^2_{m,t}=V_2$ denotes the input parameter of the second subsurrogate model with unit V.} The frequency, $f_m$, serves as an additional input parameter for the entire surrogate model. The outputs of this overall surrogate model are the multiphysics responses, denoted as $S_{11}$.
\begin{figure}[!t]
\centering
\includegraphics[width=3.5in]{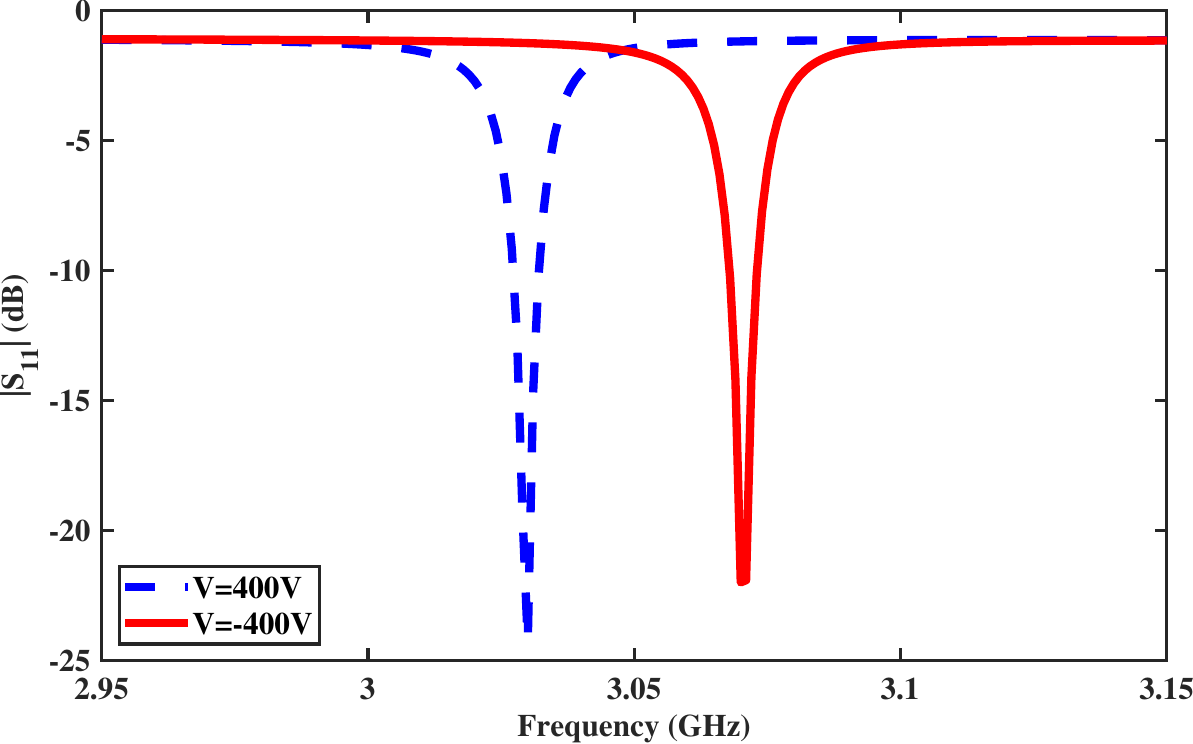}
\caption{Comparison of the magnitude of $S_{11}$ of the tunable evanescent mode-cavity filter with the same geometrical size $[{W}\;{L}\;{H}]^T = [16.5178\; 15.2459\; 243.874]^T$ [mm mm um] under different input voltages of 400 V and -400 V.} \vspace{-3.4mm}
\label{V_Po400_Ne400}
\end{figure}

\begin{figure*}[!t]
\centering
\includegraphics[width=6in]{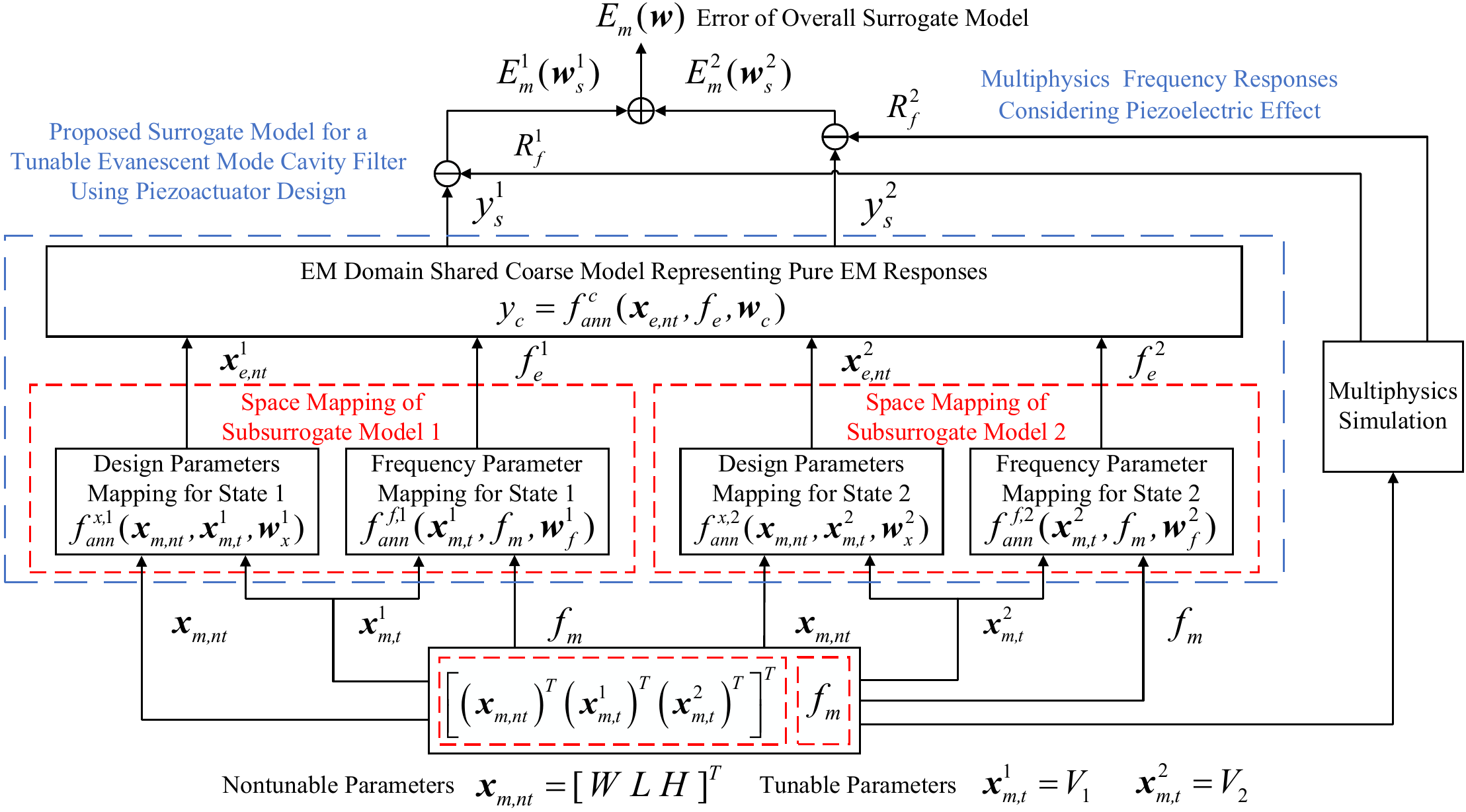}
\caption{Structure of the proposed surrogate model for the tunable evanescent-mode cavity filter.} \vspace{-3mm}
\label{Ex1_Overall_Surrogate_Model}
\end{figure*}

On the basis of preliminary experiments, the central point of the shared coarse model is designated as ${\bm{x}^k_{e,nt,c}}=[14\; 14\; 250]^T$ ($k$ = 1) for the initial optimization iteration. The trust radius is initialized as ${\bm{\delta}^k_e}=[6\;6\;60]^T$. For the $k$th optimization iteration, the region around this central point, denoted as $\bm{X}^k_e$, is defined in (\ref{Coarse_X}). A nine-level DOE is employed to generate 81 sets of EM single-physics training samples ($n_e=81$). Additionally, an eight-level DOE is employed to generate 64 sets of EM single-physics testing samples. The shared coarse model samples are generated in parallel via the ANSYS HFSS EM simulator. One ANN model is established as coarse model via the 81 training samples in NeuroModelerPlus. The shared coarse model captures the relationship between the EM single-physics $S_{11}$ and the nontunable parameters $[{W}\;{L}\;{H}]^T$ within the frequency range of 2.95 GHz--3.15 GHz. To validate the shared coarse model accuracy, 64 testing samples, which fall within the training data range but were not used during training, are used to assess the error of the trained shared coarse model. Accounting for the piezoelectric effect induced by the applied voltage, the proposed overall surrogate model represents the relationship between the multiphysics $S_{11}$ and both nontunable and tunable parameters. As illustrated in Fig. \ref{Ex1_Overall_Surrogate_Model}, our proposed overall surrogate model comprises two subsurrogate models, each of which includes one shared coarse model and two distinct mapping neural networks. The input design parameters for the first subsurrogate model are denoted as $\bm{x}_{s}^{1,k} = {[{W}\;{L}\;{H}\;{V}_1]^T}$, and those for the second subsurrogate model are denoted as $\bm{x}_{s}^{2,k} = {[{W}\;{L}\;{H}\;{V}_2]^T}$. Both subsurrogate models share the same nontunable parameters ${[{W}\;{L}\;{H}]^T}$ but have distinct tunable parameters ${V}_1$ and ${V}_2$. The initial point for the overall surrogate model is set to $\bm{x}^0_m = [14\; 14\; 250\; 0\; 0]^T$, as depicted in Fig. \ref{Ex1_Responses_of_start_and_Spec12}. During the first optimization iteration, specifically when $k$=1, we designate the central point as ${\bm{x}^k_{m,c}}={\bm{x}^0_m}=[14\; 14\; 250]^T$. In our methodology, $\bm{x}^{1,k}_{m,s,c} = [14\; 14\; 250\; 0]^T$ is identified as the central point for the first subsurrogate model, whereas $\bm{x}^{2,k}_{m,s,c} = [14\; 14\; 250\; 0]^T$ serves as the central point for the second subsurrogate model. We initialize $\bm{\delta}^k_m$ ($k$ = 1) to $[4\; 4\; 50\; 100\; 100]^T$. For the first subsurrogate model, $\bm{\delta}^{1,k}_{m,s}$ = $[4\; 4\; 50\; 100]^T$ is obtained as the trust radius $\bm{\delta}^k_m$, and for the second subsurrogate model, $\bm{\delta}^{2,k}_{m,s}$ = $[4\; 4\; 50\; 100]^T$ is similarly obtained. The region for each subsurrogate model, $\bm{X}^{i,k}_{m,s}$, around its central point $\bm{x}^{i,k}_{m,s,c}$, is delineated in (\ref{Sub_X}), whereas the overall model region, $\bm{X}^k_m$, around its central point $\bm{x}^k_{m,c}$, is defined in (\ref{Overall_X}) for the $k$th optimization iteration. Following each iteration, the trust radii for both the subsurrogate models and the overall model are updated in accordance with (\ref{r})--(\ref{delta_overall_nt}). To ensure reliable mapping between the shared coarse model and the overall surrogate model, it is imperative that the trust radius of the former exceeds that of the latter. The trust radius of the shared coarse model is updated after each iteration via equation (\ref{delta_coarse}). We employ five DOE levels to formulate the design parameter samples, generating 25 sets of multiphysics training samples, denoted as ${n_m}=25$. The testing samples comprise 20 sets of random data, which fall within the range of the training data but were not utilized during the training phase. Table~\ref{Ex1_Train_and_Testing_Samples} presents the ranges for the training samples of the EM single-physics field shared coarse model and the overall multiphysics field surrogate model. The multiphysics samples for both subsurrogate models are concurrently generated by the COMSOL MULTIPHYSICS 6.1 simulator across multiple machines with various processing cores \cite{SM13}. The overall surrogate model is constructed by concurrently developing two subsurrogate models in NeuroModelerPlus software. During this procedure, the weight parameters in the neural network mapping models within each subsurrogate model are fine-tuned until the model error falls below a user-defined threshold of 2\%, whereas the weight parameters in the shared coarse model remain unchanged.

\begin{figure}[!t]
\centering
\includegraphics[width=3.4in]{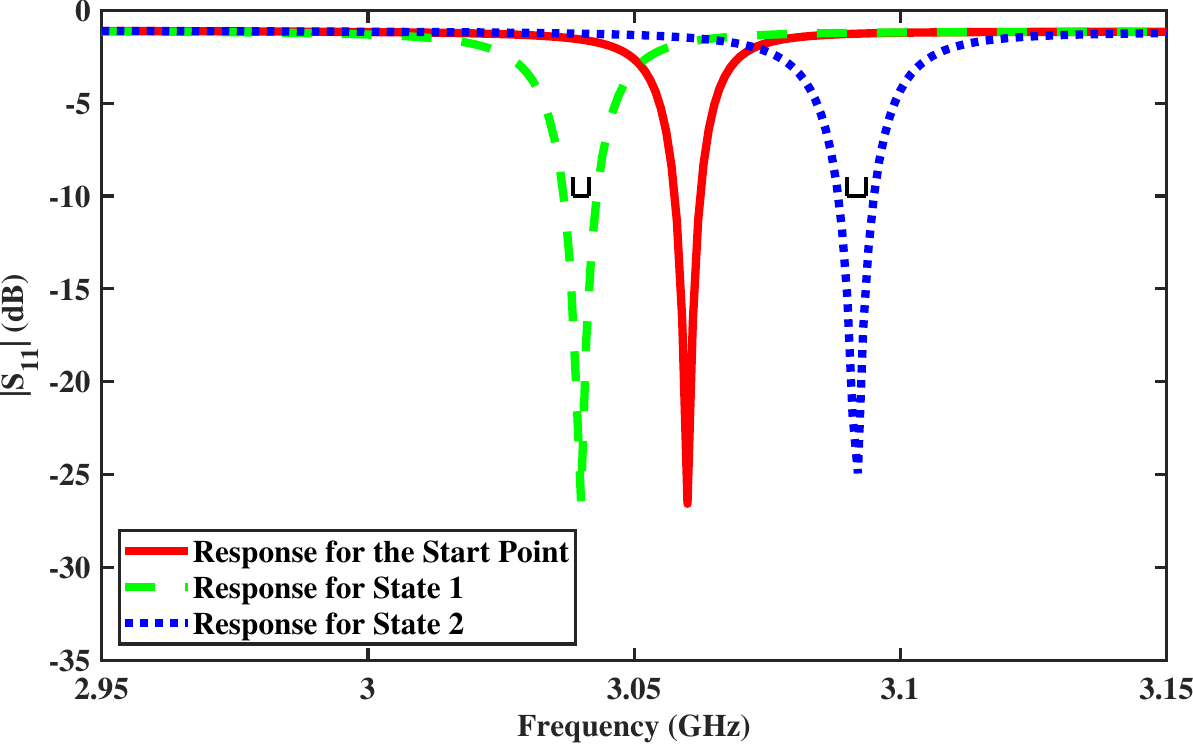}
\caption{Magnitude (in decibels) of $S_{11}$ of the multiphysics simulation for the starting point and different tuning states after 3 optimization iterations for the tunable evanescent-mode cavity filter.} \vspace{-5mm}
\label{Ex1_Responses_of_start_and_Spec12}
\end{figure}

\begin{table}[!t]
	\renewcommand{\arraystretch}{1.2}
	\caption{Training and Testing Samples of EM Field Shared Coarse Model and Multiphysics Field Overall Surrogate Model for the Tunable Evanescent-Mode Cavity Filter Example}
	\label{Ex1_Train_and_Testing_Samples}
	\centering
	\newcommand{\tabincell}[2]{\begin{tabular}{@{}#1@{}}#2\end{tabular}}
	\begin{tabular}{p{2cm}<{\centering} |p{1.4cm}<{\centering} |p{0.8cm}<{\centering} |p{0.8cm}<{\centering} |p{0.8cm}<{\centering} }
		\hline\hline
		\multicolumn{2}{c|}{\multirow{2}{*}{\tabincell{c}{Model Input Variables}}} &  \multicolumn{3}{c}{ Training Sample Range } \\
		\cline{3-5}
		\multicolumn{2}{c|}{} & Min & Max & Step \\
		\hline
		\multirow{3}{*}{\tabincell{c}{  {EM Data} \\ (Shared Coarse \\ Model)} } 
		& $W$ (mm) & 8 & 20 & 1.5  \\
		\cline{2-5}
		& $L$ (mm) & 8 & 20 & 1.5 \\
		\cline{2-5}
		& $H$ (um) & 190 & 310 & 15  \\
		\hline
		\multirow{5}{*}{\tabincell{c}{ {Multiphysics Data} \\ (Overall \\ Surrogate Model) } } 
		& $W$ (mm) & 10 & 18 & 2  \\
		\cline{2-5}
		& $L$ (mm) & 10 & 18 & 2 \\
		\cline{2-5}
		& $H$ (um) & 200 & 300 & 25  \\ 
		\cline{2-5}
		& $V_1$ (V) & -100 & 100 & 50  \\
		\cline{2-5}
		& $V_2$ (V) & -100 & 100 & 50  \\
		\hline\hline
	\end{tabular} 
\end{table}

\begin{figure}[!t]
	\centering
	\includegraphics[width=3.4in]{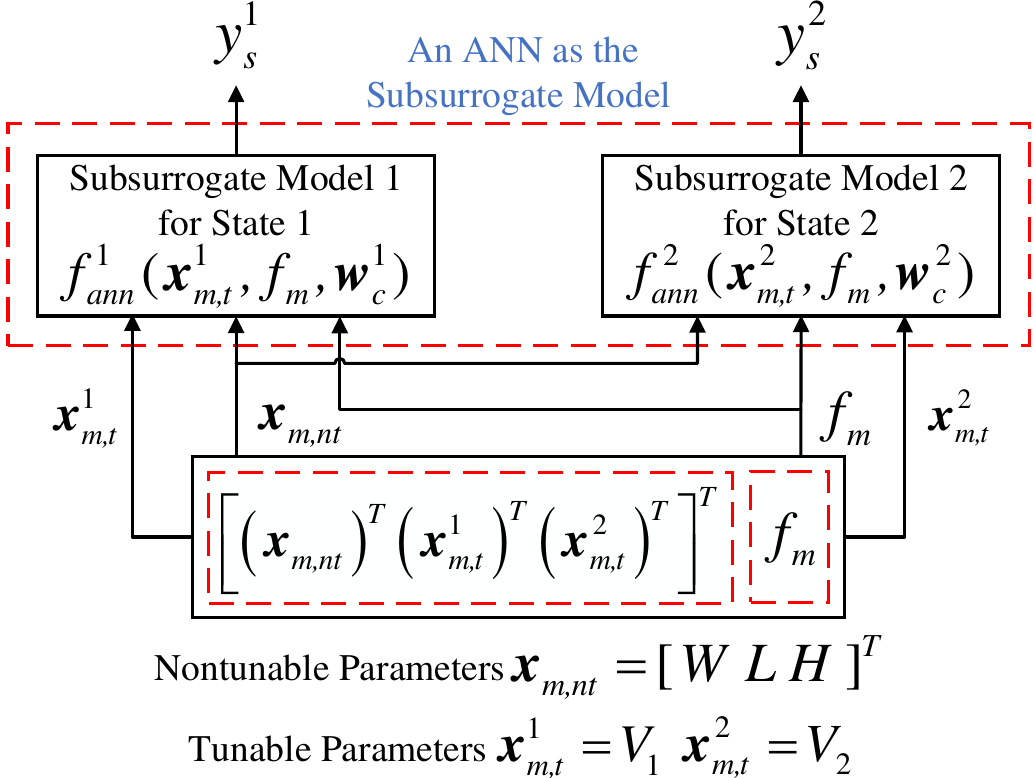}
	\caption{Structure of the surrogate model using an ANN model as the subsurrogate model for the tunable evanescent-mode cavity filter.} \vspace{-3mm}
	\label{Ex1_ANN_Surrogate_Model}
\end{figure}

\begin{figure}[!t]
\centering
\includegraphics[width=3.4in]{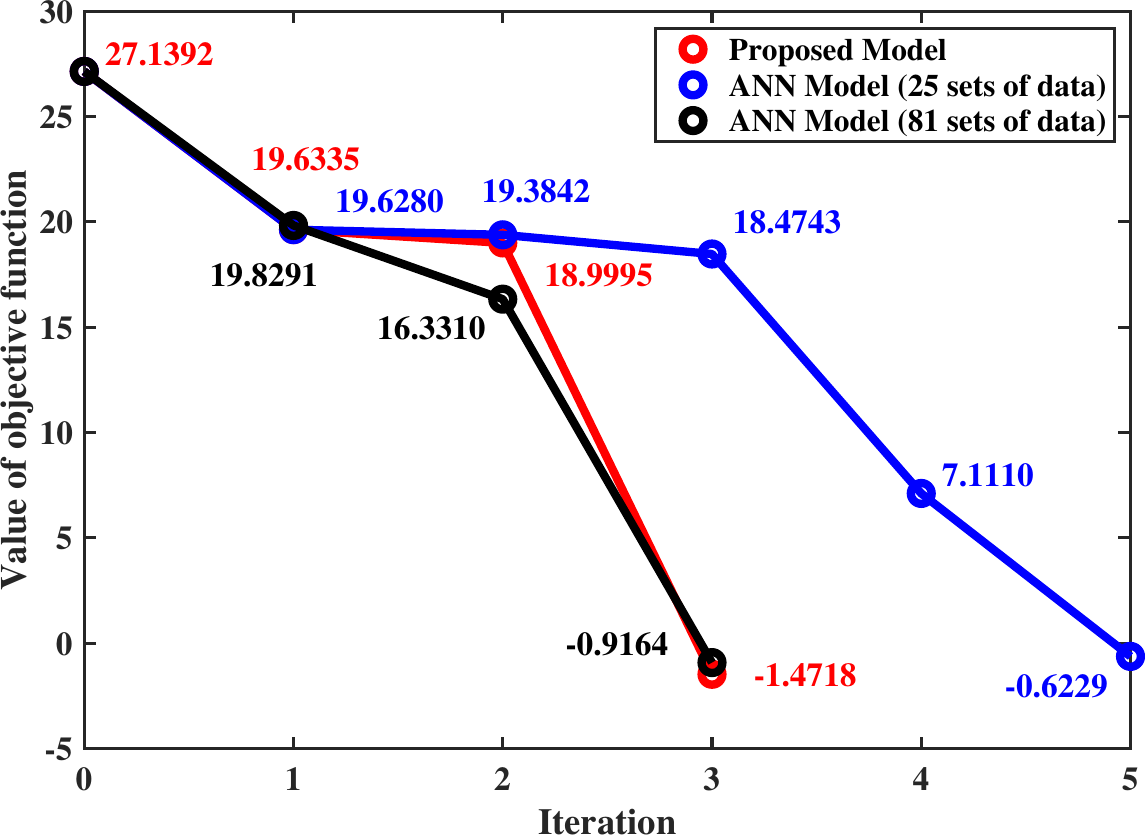}
\caption{Values of the objective function of the tunable evanescent-mode cavity filter using the proposed technique.} \vspace{-5mm}
\label{Ex1_Values_of_objective_function}
\end{figure}

In our proposed space-mapping-based surrogate-assisted multiphysics optimization technique, the values of the nontunable design parameters are constrained by both tuning states concurrently, whereas tunable design parameters are individually limited by their corresponding tuning state. After three iterations, we obtain the optimal solution \({\bm{x}^*}=[6.48726\; 23.9615\; 185.809\; 195.783\; -557.832]^T\), which satisfies two sets of tuning states within the effective range of the tunable parameter [-600V 600V]. The tunable design parameter for the first set of tuning states is $({\bm{x}^1_{m,nt}} = V_1 = 195.783)$, and that for the second set of tuning states is $({\bm{x}^2_{m,nt}} = V_2 = -557.832)$. Fig. \ref{Ex1_Responses_of_start_and_Spec12} shows the final optimization solutions obtained via our proposed space-mapping-based optimization technique.

{To illustrate the advantages of our method, we applied an existing technique to optimize the same tunable evanescent-mode cavity filter example. The overall surrogate model of this existing technique consists of two subsurrogate models, each obtained by training an ANN model in NeuroModelerPlus software with multiphysics data. The structure of the surrogate model of this technique is shown in Fig. \ref{Ex1_ANN_Surrogate_Model}.} The design parameters of the overall surrogate model are represented as \(\bm{x}_m = [W\;L\;H\;V_1\;V_2]^T\). Both subsurrogate models share identical nontunable design parameters, denoted as \(\bm{x}_{m,nt} = [W\;L\;H]^T\), which are simultaneously constrained by two tuning states during the optimization iterations. For the first subsurrogate model, its tunable design parameter \(V_1\) is limited by the first tuning state, whereas for the second subsurrogate model, its tunable design parameter \(V_2\) is governed by the second tuning state. The initial design parameters are set as \(\bm{x}^0_m = [14\; 14\; 250\; 0\; 0]^T\). The multiphysics response, with a frequency range of 3 GHz to 3.1 GHz, is illustrated in Fig. \ref{Ex1_Responses_of_start_and_Spec12}. {In order to make a more complete comparison, we conduct two sets of comparative experiments. We train the subsurrogate model in existing techniques with 25 sets of multiphysics data and 81 sets of multiphysics data respectively.} After 5 optimization iterations, the optimal solution ${\bm{x}^*} = [6.47613\; 27.3691\; 249.51\; 458.437\; -441.727]^T$ of the overall surrogate model trained with 25 sets of samples is obtained. After 3 optimization iterations, the optimal solution ${\bm{x}^*} = [6.47613\; 27.3691\; 249.51\; 458.437\; -441.727]^T$ of the overall surrogate model trained with 81 sets of samples is obtained. Table \ref{Ex1_Comparisons} presents a comparative analysis of the technique proposed in this article and two existing techniques for designing the tunable evanescent mode cavity filter {\cite{Tunable_Filter6}}. {Total time is composed of the EM evaluation time, multiphysics  evaluation simulation time, surrogate model training and optimization time in each iteration.} {Parallel computation scheme is used to perform EM simulation for generating the EM single-physics training samples.} Regardless of whether the existing technique employs 81 or 25 sets of training samples, our proposed method yields a more accurate overall surrogate model. {The proposed method reduces the total consumption time by 27.3\% compared to the method using ANN model trained by 25 sets of multiphysis samples, and 46.8\% compared to the method using ANN model trained by 81 sets of multiphysis samples.} Furthermore, our optimization approach converges to the optimal solution in fewer iterations, utilizing only 25 sets of training samples. The objective function values across all iterations for both the proposed technique and the two existing methods are depicted in Fig. \ref{Ex1_Values_of_objective_function}. Our proposed optimization converges within 3 iterations, achieving the optimal solution for all the tuning states with fewer iterations than the existing techniques do. Consequently, our proposed method demonstrates superior efficiency relative to current techniques.
\begin{table}[!t]
	\renewcommand{\arraystretch}{1.5}
	\centering
	\caption{Comparison of Different Optimization Techniques for the Tunable Evanescent-Mode Cavity Filter Example}
	\label{Ex1_Comparisons}
	\begin{tabular}{>{\centering\arraybackslash}m{2.3cm}
			|>{\centering\arraybackslash}m{1.5cm}
			|>{\centering\arraybackslash}m{1.5cm}
			|>{\centering\arraybackslash}m{1.5cm}}
		\hline\hline
		Training Model  % 移除了加粗
		& \multicolumn{2}{c|}{\makecell{ANN Model as the \\ Subsurrogate Model [37]}}  % 移除了加粗
		& Proposed Model \\ % 移除了加粗
		\hline
		No. of EM Samples per Iteration & 0 & 0 & $81 \times 2$ \\
		\hline
		No. of MP Samples per Iteration & $25 \times 2$ & $81 \times 2$ & $25 \times 2$ \\
		\hline
		No. of Iterations & 5 & {3} & 3 \\
		\hline
		EM Evaluation Time & -- & -- & $32\ \text{min} \times 3$ \\
		\hline
		Multiphysics Evaluation Time & $157\ \text{min} \times 5$ & $361\ \text{min} \times 3$ & $157 \ \text{min} \times 3$ \\
		\hline
		Surrogate Model Training and Optimization Timen & $2\ \text{min} \times 5$ & $2\ \text{min} \times 3$ & $4\ \text{min} \times 3$ \\
		\hline
	    Total Time & 13.25 h & 18.15 h & 9.65 h \\
		\hline
		Total Time Saving & -- & -- & {21.6\%\textuparrow VS ANN(DOE25) 49.7\%\textuparrow VS ANN(DOE81)} \\
		\hline\hline
	\end{tabular}
    \par\vspace{1mm} % 添加垂直间距
    \begin{minipage}{\textwidth}
	\raggedright % 左对齐
	\small % 小号字体
	\hangindent=1.5em % 首行缩进
	
	$*$ EM and multiphysics data are generated in parallel.
	
%	\vspace{4pt} % 段落间距
	$\dagger$ Surrogate model training and optimization time with parallel \\ computation technique.
    \end{minipage}
\end{table}

\subsection{Multiphysics Optimization of a Tunable Four-Pole Waveguide Filter Using a Piezoactuator}
The second example is a tunable four-pole waveguide filter \cite{Neuro_TF_Opt}, as shown in Fig.~\ref{Ex2_Structure}. This filter has a standard WR-75 waveguide cross-section with a width of 19.05 mm and a height of 9.525 mm \cite{Ex2}. The length of the filter is 77.6 mm, and the thickness of all coupling windows is 2 mm. $h_1$ and $h_2$ are the heights of the tuning posts in the coupling windows, and $h_{c1}$ and $h_{c2}$ are the heights of the tuning posts in the center of the cavities. The nontunable design parameters for the overall surrogate model are $h_1$, $h_2$, $h_{c1}$, and $h_{c2}$. $V_1$ and $V_2$ are the electronic potentials applied to the piezoactuator, which are multiphysics design parameters. These voltages can cause deformation and displacement of the piezoactuators, which changes the gaps between the bottom side of the piezoactuators and the top of the tuning posts in the center of the cavities, thus affecting the response of the tunable four-pole waveguide filter. The tunable design parameters for the overall surrogate model are $V_1$ and $V_2$, whose ranges are from -600 V to 600 V. Fig. \ref{Ex2_Struc_one_point} shows the deformed structure of the tunable four-pole waveguide filter when $[h_1\;h_2\;h_{c1}\;h_{c2}]^T = [3.51967\; 4.31315\; 3.15755\; 2.90962]^T$ [mm mm mm mm] and $[V_1\;V_2]^T = [100\; -100]^T$ [V V]. The positive voltage deflects the piezoactuator downward, whereas the negative voltage deflects it upward. For the tunable four-pole waveguide filter example, we have three sets of tuning states:
\begin{equation}
\begin{array}{l}
 \text{State 1:}\left| {{S_{11}}} \right| \le -20 \;\text{dB}, \;10.75 \;\text{GHz} \le {{f}} \le 11.05 \;\text{GHz} \vspace{1mm}\\
 \text{State 2:}\left| {{S_{11}}} \right| \le -23.5 \;\text{dB}, \;10.85 \;\text{GHz} \le {{f}} \le 11.15 \;\text{GHz} \vspace{1mm} \\
 \text{State 3:}\left| {{S_{11}}} \right| \le -22.5 \;\text{dB}, \;10.95 \;\text{GHz} \le {{f}} \le 11.25 \;\text{GHz}
\end{array}
\label{Ex2_tuning_states}
\end{equation}

\begin{figure}[!t]
\centering
\includegraphics[width=3.4in]{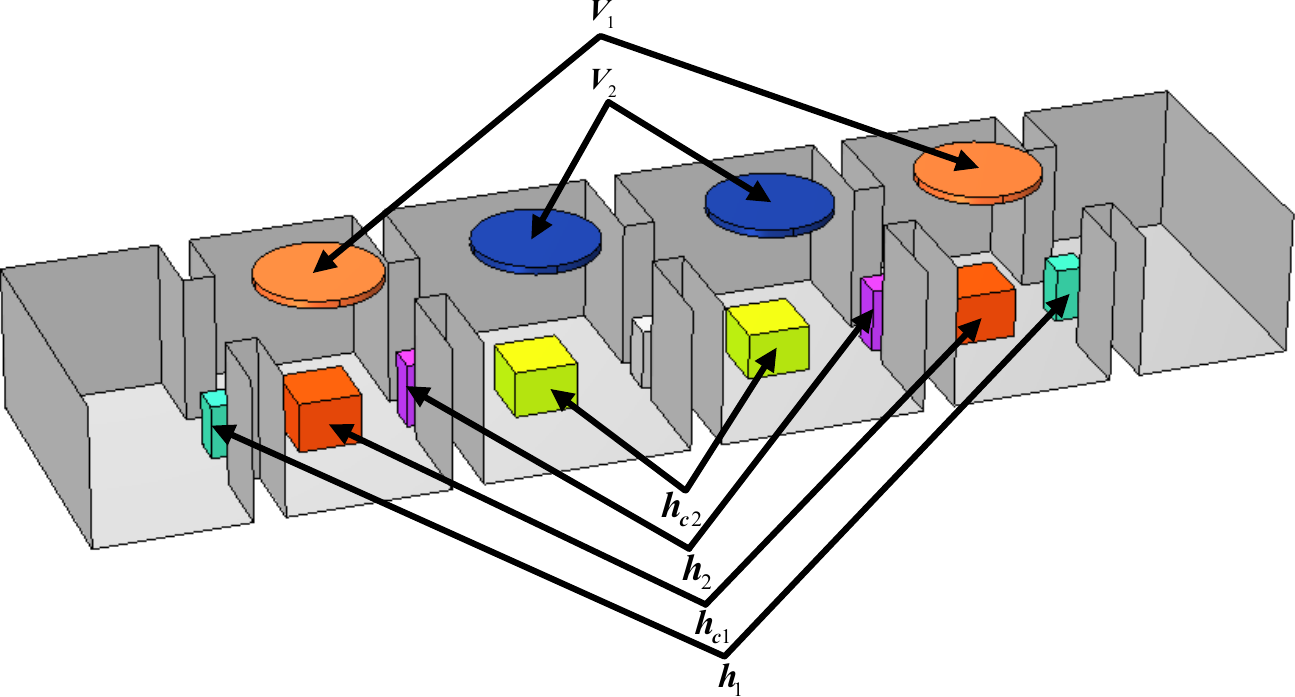}
\caption{Structure of the tunable four-pole waveguide filter using a piezoactuator.} \vspace{-3mm}
\label{Ex2_Structure}
\end{figure}

\begin{figure}[!t]
\centering
\includegraphics[width=3.5in]{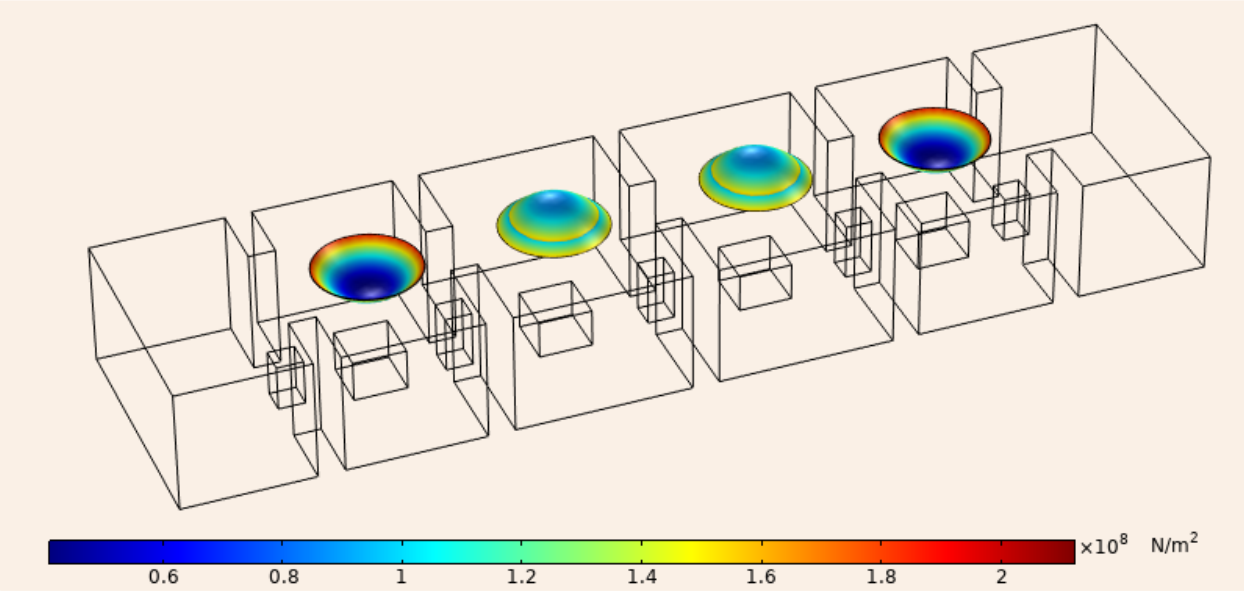}
\caption{Deformed structure of the tunable four-pole waveguide filter caused by the input voltages.} \vspace{-3mm}
\label{Ex2_Struc_one_point}
\end{figure}

\begin{figure*}[!t]
\centering
\includegraphics[width=7.3in]{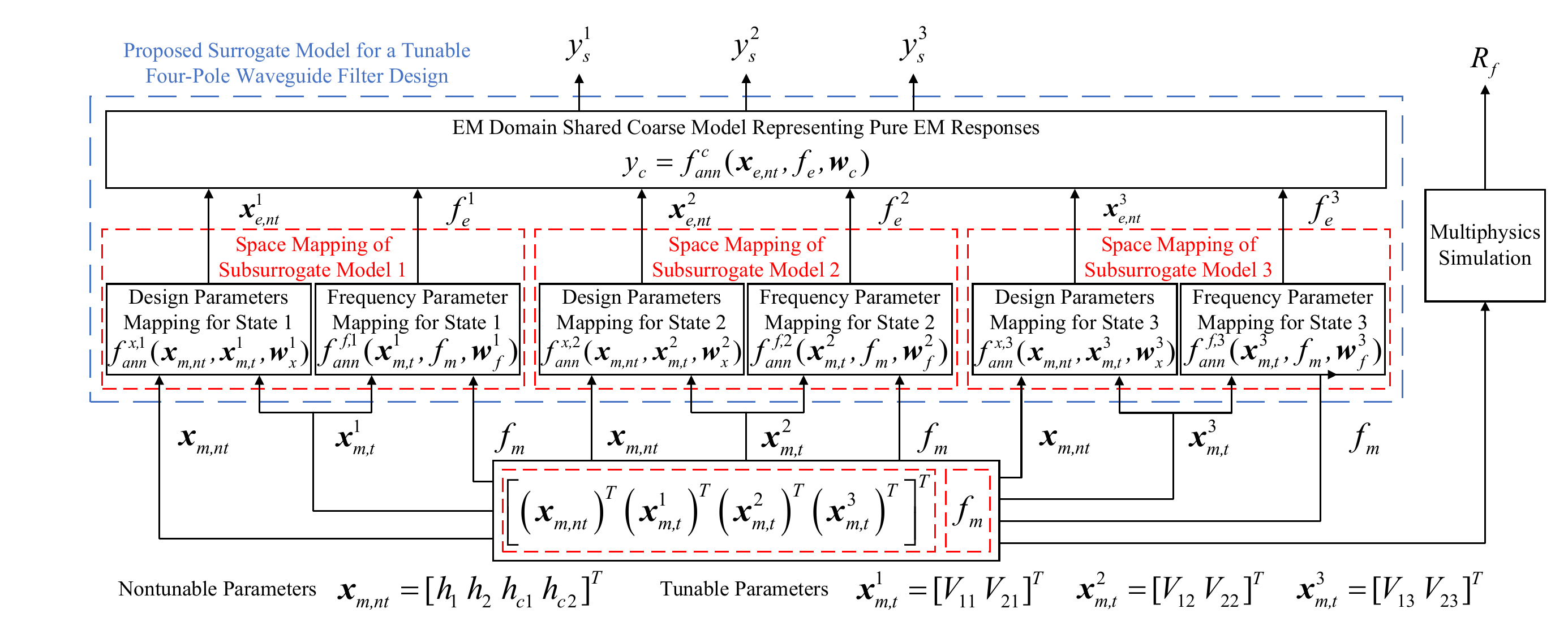}
\caption{Structure of the proposed surrogate model for this tunable four-pole waveguide filter} \vspace{-3mm}
\label{Ex2_Overall_Surrogate_Model}
\end{figure*}

\begin{figure}[!t]
\centering
\includegraphics[width=3.5in]{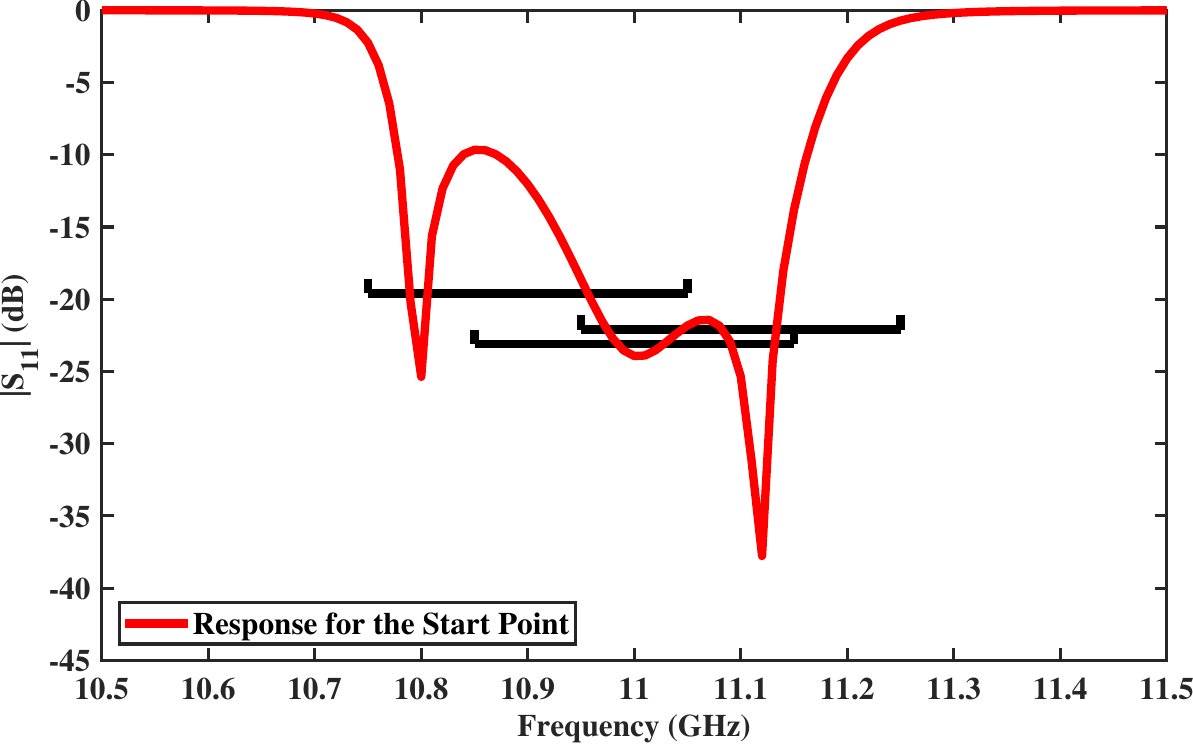}
\caption{Magnitude (in decibels) of $S_{11}$ of the multiphysics simulation of the starting point for the tunable four-pole waveguide filter.} \vspace{-3mm}
\label{Ex2_Responses_of_start}
\end{figure}

With our proposed technique, it is suggested that the overall surrogate model comprises three subsurrogate models, each corresponding to one of the three sets of tuning states. As illustrated in Fig.~\ref{Ex2_Overall_Surrogate_Model}, all the subsurrogate models share a shared coarse model and two distinct mapping neural networks. The shared coarse model captures the relationship between the EM single-physics (EM only) responses and the nontunable parameters. Consequently, the input parameters $\bm{x}^1_{e,nt}$, $\bm{x}^2_{e,nt}$, and $\bm{x}^3_{e,nt}$ for the shared coarse model are uniformly represented as $[{h_1}\;{h_2}\;{h_{c1}}\;{h_{c2}}]^T$. Additionally, the frequencies $f^1_e$, $f^2_e$, and $f^3_e$ serve as auxiliary input parameters for the shared coarse model. The output parameter for the shared coarse model is $S_{11}$. 

{The purpose of this article is multiple filters design simultaneously, each subsurrogate model represents one filter design. Therefore, the values of tuning parameters are different for different subsurrogate models (i.e., different filters). We design three filters simultaneously. There are two voltages $V_1$ and $V_2$ to be the tunable parameters for each filter. ${V_{11} \; V_{21} \; V_{12} \; V_{22} \; V_{13} \; V_{23}}$ are the controlling voltages for the first filters, second filter, and third filter respectively. }The design parameters for the entire surrogate model are defined as
\begin{equation}
	\bm{x}_m = [h_1 \; h_2 \; h_{c1} \; h_{c2} \; V_{11} \; V_{21} \; V_{12} \; V_{22} \; V_{13} \; V_{23}]^T
\end{equation}
{where $\bm{x}_{m,nt} = [h_1 \; h_2 \; h_{c1} \; h_{c2}]^T$ is concurrently constrained by three distinct sets of tuning states, and the unit is [mm mm mm mm]. The variables $\bm{x}^1_{m,t} = [V_{11} \; V_{21}]^T$, $\bm{x}^2_{m,t}=[V_{12} \; V_{22}]^T$, and $\bm{x}^3_{m,t}=[V_{13} \; V_{23}]^T$ denotes the input parameters for three subsurrogate models respectively, all in units of [V V].} The input parameters for the first subsurrogate model are denoted as $\bm{x}^1_{m,s}=[h_1 \; h_2 \; h_{c1} \; h_{c2} \; V_{11} \; V_{21}]^T$. Similarly, $\bm{x}^2_{m,s}=[h_1 \; h_2 \; h_{c1} \; h_{c2} \; V_{12} \; V_{22}]^T$ represents the input parameters for the second subsurrogate model, and $\bm{x}^3_{m,s}=[h_1 \; h_2 \; h_{c1} \; h_{c2} \; V_{13} \; V_{23}]^T$ represents the input parameters fo the third subsurrogate model. Additionally, $f_m$ serves as an auxiliary input parameter for the overall surrogate model. The output of this overarching model is characterized by the multiphysics response, denoted as $S_{11}$.

 {The initial values of the starting points are mainly determined based on the historical design experience of similar filters and structural symmetry. With reference to the geometric proportions of the standard WR-75 waveguide and the coupling structure size of similar four-pole filters, the coupling window height and the resonant cavity center column height are usually preliminarily set according to the waveguide wavelength and coupling coefficient empirical formula. We set the initial $[h_1 \; h_2 \; h_{c1} \; h_{c2}]^T=[3.4 4.08 3.26 2.96]^T$. When there is no excitation, the filter is in the baseline state, which is convenient for subsequently introducing frequency offset by adjusting the voltage. Therefore, the initial value of the adjustable parameter (voltage) is set to 0, corresponding to the state where no excitation is applied to the piezoelectric actuator. And when the voltage is set to 0, it simplifies the structure of the four-pole waveguide filter and reduces the complexity of the parameter space.} The starting point for the overall surrogate model is set as \(\bm{x}^0_{m,c} = [3.4\; 4.08\; 3.26\; 2.96\; 0\; 0\; 0\; 0\; 0\; 0]^T\), as illustrated in Fig. \ref{Ex2_Responses_of_start}. {In our proposed method, even if the initial value responds poorly, the proposed multistate tuning-drive optimization technique is still effective. When the initial point is poor, using our proposed method will iterate more times, but it still takes less time than the previous methods. We can also add a pole-zero-based transfer function to address this issue.} For the first optimization iteration, denoted as \(k=1\), the center point is defined as \({\bm{x}^k_{m,c}}={\bm{x}^0_m}\). The trust radius \(\bm{\delta}^k_m\) for \(k=1\) is initialized to \( [0.08\; 0.08\; 0.08\; 0.08\; 50\; 50\; 50\; 50\; 50\; 50 ]^T\). The center point and trust radius for each subsurrogate model are derived from \({\bm{x}^k_{m,c}}\) and \(\bm{\delta}^k_m\), respectively. For example, the center point of the first subsurrogate model is  \(\bm{x}^{1,k}_{m, s,c} = [3.4\; 4.08\; 3.26\; 2.96\; 0\; 0]^T\) and the trust radius is \(\bm{\delta}^{1,k}_{m,s} = [0.08\; 0.08\; 0.08\; 0.08\; 50\; 50]^T\). The second and third subsurrogate models have identical values for center point and trust radius. The training regions for both the subsurrogate model and the overall model are delineated by these center points and trust radii, as described in formulas (\ref{Sub_X}) and (\ref{Overall_X}). Following each optimization iteration, the trust radii for both the subsurrogate model and the overall models are updated according to Equations (\ref{r})--(\ref{delta_overall_nt}). In each optimization iteration, the center point of the shared coarse model is derived from ${\bm{x}^k_{m,c}}$. For the initial iteration, denoted as $k$=1, this center point is set to $\bm{x}^k_{e,nt,c}=[3.4\; 4.08\; 3.26\; 2.96]^T$. To ensure reliable mapping between the shared coarse model and the overall surrogate model, the trust radius of the shared coarse model must exceed that of the overall surrogate model. The trust radius of the shared coarse model is updated after each iteration via (\ref{delta_coarse}). Consequently, the trust radius of the shared coarse model is set to ${\bm{\delta}^k_e}=[0.1\; 0.1\; 0.1\; 0.1]^T$. The region of the shared coarse model, $\bm{X}^k_e$, around the central point for the $k$th optimization iteration is defined in (\ref{Coarse_X}). The DOE sampling method is employed to generate design parameter samples. For the development of the shared coarse model, we utilize nine DOE levels to generate training samples, resulting in a total of 81 sets of training samples (i.e., ${n_e}=81$), and eight DOE levels to generate testing samples, yielding a total of 64 sets of testing samples. For the development of the subsurrogate models, we employ only five DOE levels to generate training samples, resulting in a total of 25 sets of training samples (i.e., ${n_m}=25$). The testing samples consisted of 20 sets of random data. Table~\ref{Ex2_Train_and_Testing_Samples} presents the ranges of the training samples for the EM single-physics (EM only) field shared coarse model and the overall multiphysics field surrogate model. We use a parallel technique to generate shared coarse model samples for the EM single-physics field in the ANSYS HFSS EM simulator and subsurrogate model samples in the COMSOL MULTIPHYSICS 6.1 simulator for the multiphysics field. A total of 81 sets of EM single-physics domiain training samples are used to establish an ANN model in NeuroModerlerPlus software. The model serve as the shared coarse model. During the training phase, the weight parameters of the ANN model are meticulously optimized. Sixty-four sets of testing samples are employed to assess the model accuracy. In the subsequent stages of this iterative optimization process, the weight parameters from the trained shared coarse model remain fixed. The overall surrogate model is constructed by integrating three subsurrogate models, which are developed by optimizing the weight parameters in the mapping neural network models via 25 sets of multiphysics training samples. To evaluate the efficacy of the overall surrogate model, we utilize 20 sets of random testing samples. These samples fall within the range of the training data but were not used during the training process. Training of the overall surrogate model concludes once the model error meets or falls below a predetermined threshold, set at 2\%.
\begin{table}[!t]
	\renewcommand{\arraystretch}{1.2}
	\caption{Training and Testing Samples of EM ield Shared Coarse Model and Multiphysics filed Overall Surrogate Model for the Tunable Four-Pole Waveguide Filter Example}
	\label{Ex2_Train_and_Testing_Samples}
	\centering
	\newcommand{\tabincell}[2]{\begin{tabular}{@{}#1@{}}#2\end{tabular}}
	\begin{tabular}{p{2cm}<{\centering} |p{1.4cm}<{\centering} |p{0.8cm}<{\centering} |p{0.8cm}<{\centering} |p{0.8cm}<{\centering} }
		\hline\hline
		\multicolumn{2}{c|}{\multirow{2}{*}{\tabincell{c}{Model Input Variables}}} &  \multicolumn{3}{c}{ Training Sample Range } \\
		\cline{3-5}
		\multicolumn{2}{c|}{} & Min & Max & Step \\
		\hline
		\multirow{4}{*}{\tabincell{c}{  EM Data \\ (Shared Coarse \\ Model)} } 
		& $h_1$ (mm) & 3.30 & 3.50 & 0.025  \\
		\cline{2-5}
		& $h_2$ (mm) & 3.98 & 4.18 & 0.025 \\
		\cline{2-5}
		& $h_{c1}$ (mm) & 3.16 & 3.36 & 0.025  \\
		\cline{2-5}
		& $h_{c2}$ (mm) & 2.86 & 3.06 & 0.025  \\
		\hline
		\multirow{9}{*}{\tabincell{c}{ Multiphysics Data \\ (Overall Surrogate \\ Model) } } 
		& $h_1$ (mm) & 3.32 & 3.48 & 0.04  \\
		\cline{2-5}
		& $h_2$ (mm) & 4.00 & 4.16 & 0.04 \\
		\cline{2-5}
		& $h_{c1}$ (mm) & 3.18 & 3.34 & 0.04  \\ 
		\cline{2-5}
		& $h_{c2}$ (mm) & 2.88 & 3.04 & 0.04  \\
		\cline{2-5}
		& $V_{11}$ (V) & -100 & 100 & 50  \\
		\cline{2-5}
		& $V_{21}$ (V) & -100 & 100 & 50  \\
		\cline{2-5}
		& $V_{12}$ (V) & -100 & 100 & 50  \\
		\cline{2-5}
		& $V_{22}$ (V) & -100 & 100 & 50  \\
		\cline{2-5}
		& $V_{13}$ (V) & -100 & 100 & 50  \\
		\cline{2-5}
		& $V_{23}$ (V) & -100 & 100 & 50  \\
		\hline\hline
	\end{tabular}
\end{table}

\begin{figure}[!t]
\centering
\subfigure[]{\includegraphics[width=3.4in]{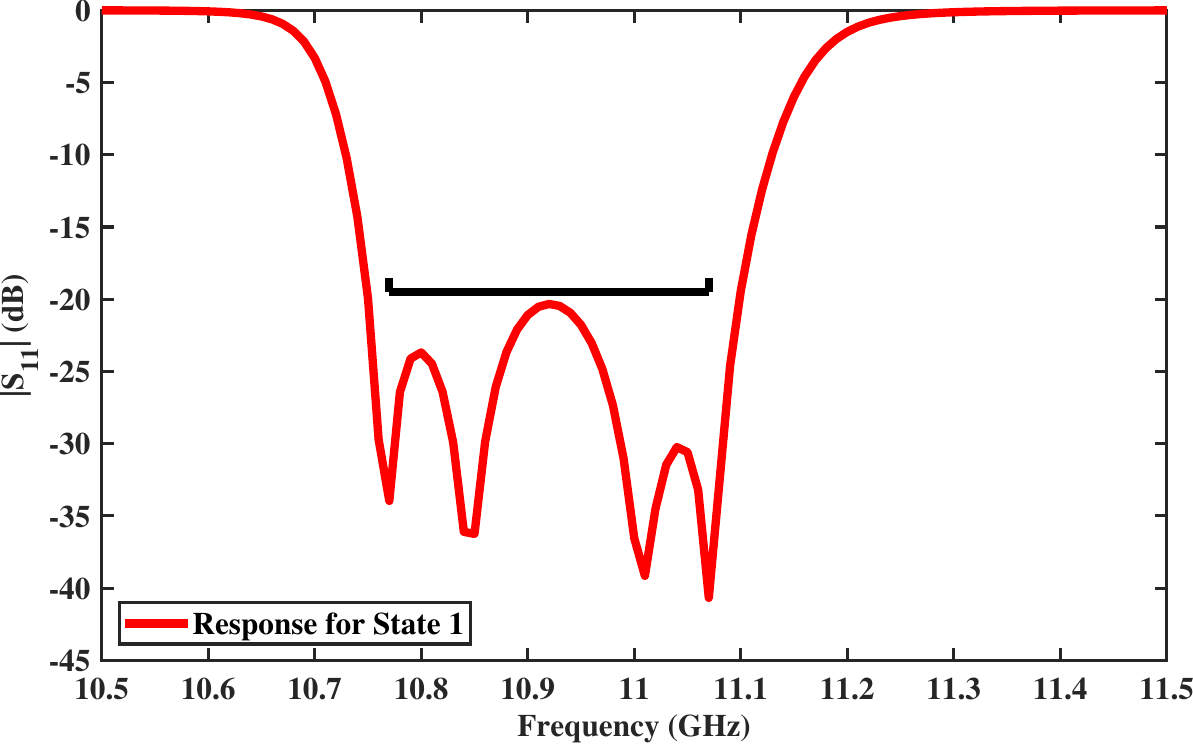}}
\subfigure[]{\includegraphics[width=3.4in]{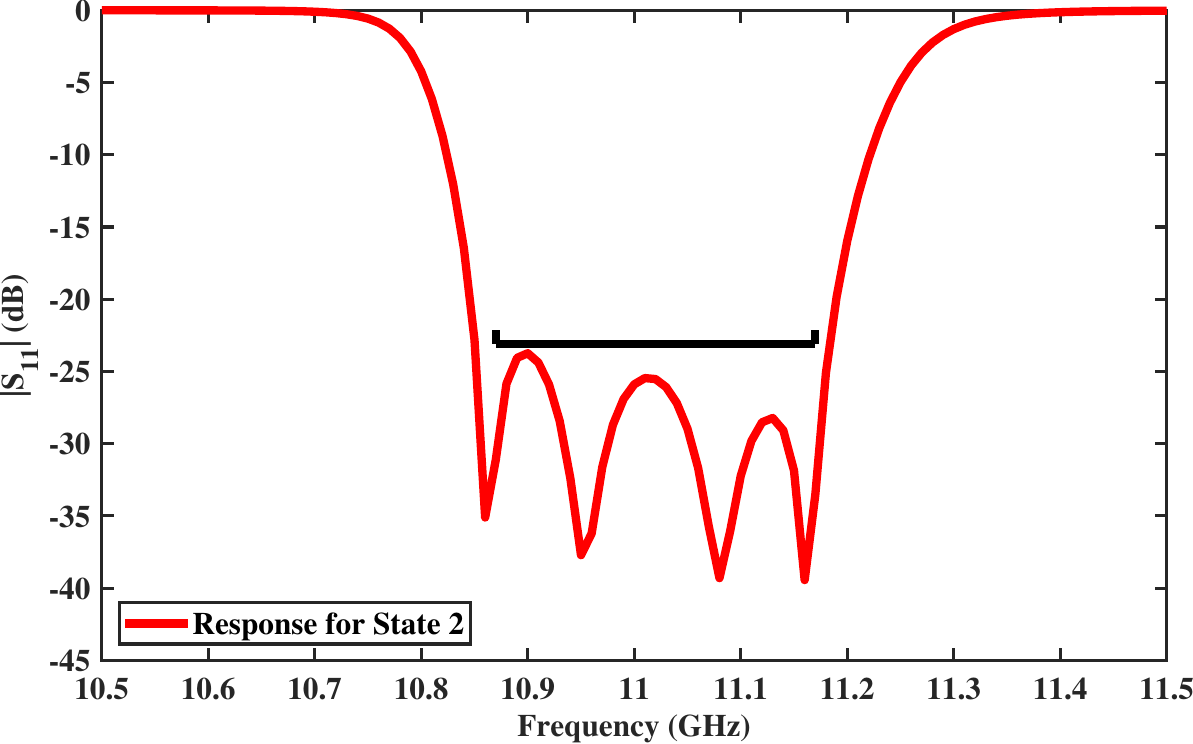}}
\subfigure[]{\includegraphics[width=3.4in]{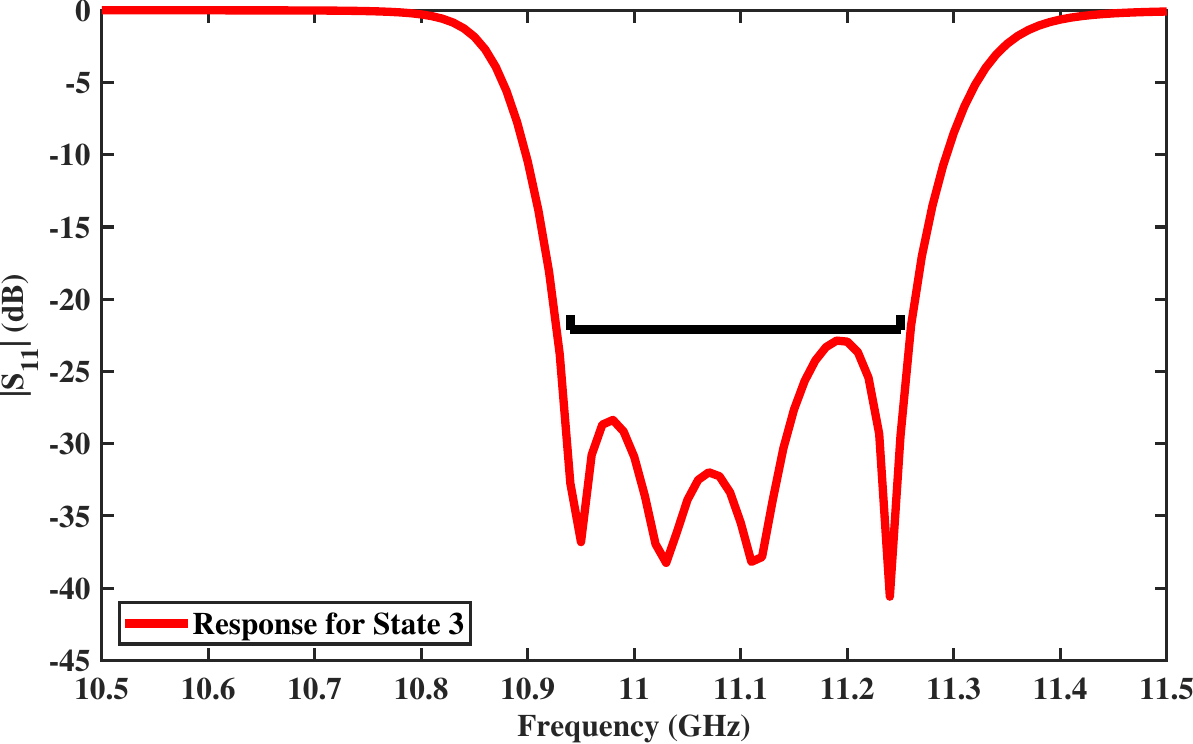}}
\caption{Magnitude (in decibels) of $S_{11}$ via the multiphysics simulation under different design specifications after 5 optimization iterations for the tunable four-pole waveguide filter example. (a) Multiphysics response at the optimal solution for state 1. (b) Multiphysics response at the optimal solution for state 2. (c) Multiphysics response at the optimal solution for state 3.} \vspace{-5mm}
\label{Ex2_Responses_of_Spec123}
\end{figure}

The proposed tunable microwave filter optimization technique allows convergence to the optimal solution across all tuning states within only 5 iterations. The final optimal solution is represented as ${\bm{x}^*}=[3.53303\; 4.31909\; 3.15955\; 2.91159\; 439.578\; 312.484\; 91.5983\ -60.076\; -310.487\; -593.379]^T$. The nontunable design parameters are denoted as ${\bm{x}_{m,nt}} = [3.53303\; 4.31909\; 3.15955\; 2.91159]^T$. The tunable design parameters for the first set of tuning states are ${\bm{x}_{m,t}^1}={[V_{11}\; V_{21}]}^T=[439.578\; 312.484]^T$. The second set of tuning states is ${\bm{x}_{m,t}^2}={[V_{12}\; V_{22}]}^T=[91.5983\ -60.076]^T$, and the third set is ${\bm{x}_{m,t}^3}={[V_{13}\; V_{23}]}^T=[-310.487\; -593.379]^T$. Fig. \ref{Ex2_Responses_of_Spec123} (a)-(c) shows the multiphysics responses of the optimal solutions that meet the three sets of tuning states. The values of the nontunable design parameters are constrained by the set of all tuning states, whereas the tunable design parameters are restricted by the corresponding tuning state such that all frequency responses satisfy the tuning state.

\begin{figure}[!t]
	\centering
	\includegraphics[width=3.4in]{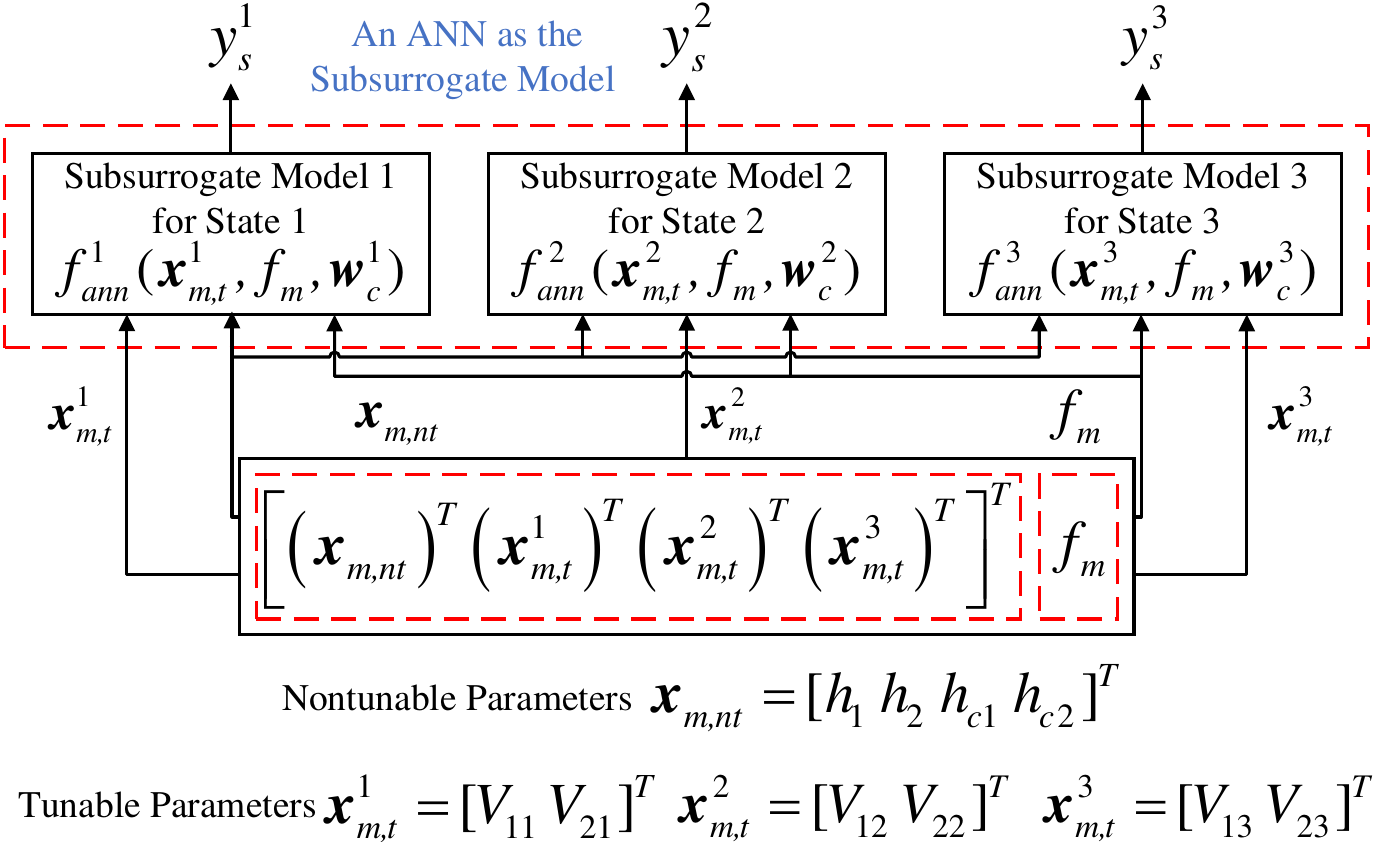}
	\caption{Structure of the surrogate model using an ANN model as the subsurrogate model for the tunable four-pole waveguide filter.} \vspace{-3mm}
	\label{Ex2_ANN_Surrogate_Model}
\end{figure}

\begin{figure}[!t]
	\centering
	\includegraphics[width=3.4in]{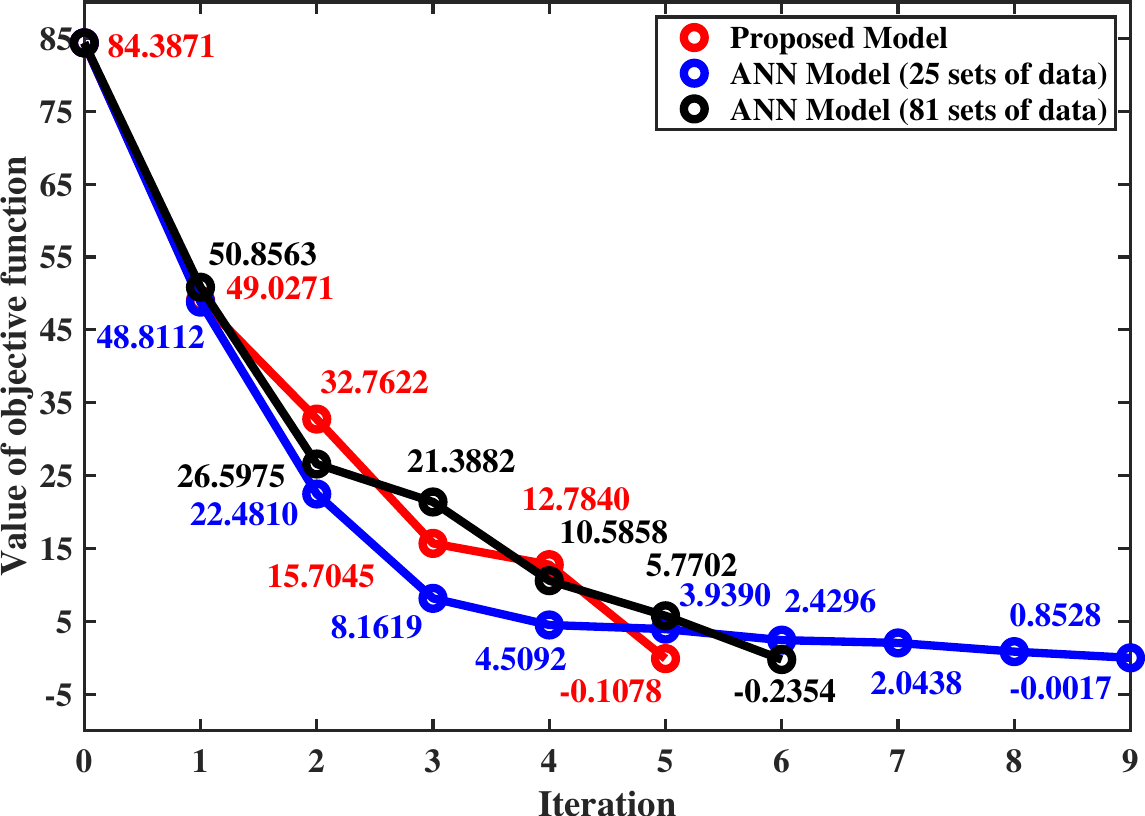}
	\caption{Values of the objective function of the tunable four-pole waveguide filter using the proposed technique.} \vspace{-5mm}
	\label{Ex2_Values_of_objective_function}
\end{figure}

\begin{table}[!t]
	\renewcommand{\arraystretch}{1.5}
	\centering
	\caption{Comparison of Different Optimization Techniques for the Tunable Four-Pole Waveguide Filter Example}
	\label{Ex2_Comparisons}
	\begin{tabular}{>{\centering\arraybackslash}m{2.3cm}
			|>{\centering\arraybackslash}m{1.5cm}
			|>{\centering\arraybackslash}m{1.5cm}
			|>{\centering\arraybackslash}m{1.5cm}}
		\hline\hline
		Training Model  % 移除了加粗
		& \multicolumn{2}{c|}{\makecell{ANN Model as the \\ Subsurrogate Model [37]}}  % 移除了加粗
		& Proposed Model \\ % 移除了加粗
		\hline
		No. of EM Samples per Iteration & 0 & 0 & $81 \times 3$ \\
		\hline
		No. of MP Samples per Iteration & $25 \times 3$ & $81 \times 3$ & $25 \times 3$ \\
		\hline
		No. of Iterations & 8 & 6 & 5 \\
		\hline
		EM Evaluation Time & -- & -- & $72\ \text{min} \times 5$ \\
		\hline
		Multiphysics Evaluation Time & $289\ \text{min} \times 8$ & $603\ \text{min} \times 6$ & $289 \ \text{min} \times 5$ \\
		\hline
		Surrogate Model Training and Optimization Timen & $3\ \text{min} \times 8$ & $3\ \text{min} \times 6$ & $5\ \text{min} \times 5$ \\
		\hline
		Total Time & 38.9 h & 60.6 h & 30.5 h \\
		\hline
		Total Time Saving & -- & -- & {27.3\%\textuparrow VS ANN(DOE25) 46.8\%\textuparrow VS ANN(DOE81)} \\
		\hline\hline
	\end{tabular}
	\par\vspace{1mm} % 添加垂直间距
	\begin{minipage}{\textwidth}
		\raggedright % 左对齐
		\small % 小号字体
		\hangindent=1.5em % 首行缩进
		
		$*$ EM and multiphysics data are generated in parallel.
		
		%	\vspace{4pt} % 段落间距
		$\dagger$ Surrogate model training and optimization time with parallel \\ computation technique.
	\end{minipage}
\end{table}

For comparative purposes, we employed an existing technique to optimize the tunable four-pole waveguide filter example. Importantly, the subsurrogate models in this technique differ from those proposed in our article. {This existing method constructs the overall surrogate model by integrating three subsurrogate models, each derived from ANN training on multiphysics data using the NeuroModelerPlus software, as depicted in Fig. \ref{Ex2_ANN_Surrogate_Model}.} An ANN model was directly trained to capture the relationship between multiphysics frequency responses and design parameters, serving as a subsurrogate model. The design parameters of the overall surrogate model are represented by \(\bm{x}_m = [h_1\;h_2\;h_{c1}\;h_{c2}\;V_{11}\;V_{21}\;V_{12}\;V_{22}\;V_{13}\;V_{23}]^T\). All three subsurrogate models share identical nontunable design parameters, denoted as \(\bm{x}_{m,nt} = [h_1\;h_2\;h_{c1}\;h_{c2}]^T\), which are concurrently constrained by three tuning states throughout the optimization iteration process. The tunable design parameter \([V_{11}\;V_{21}]^T\) of the first subsurrogate model is governed by the first tuning state set. Similarly, the tunable design parameter \([V_{12}\;V_{22}]^T\) of the second subsurrogate model adheres to the second tuning state set, whereas the tunable design parameter \([V_{13}\;V_{23}]^T\) of the third subsurrogate model complies with the third tuning state set. The initial values are set to \(\bm{x}^0 = [3.2 \;3.84\; 3.02\; 2.72\; 0\; 0\; 0\; 0\; 0\; 0]^T\), as illustrated in Fig. \ref{Ex2_Responses_of_start}. To ensure experimental rigor, an orthogonal distribution was utilized to generate multiple samples. {To facilitate a more complete comparison, we perform two comparative experiments by training the subsurrogate models in the existing approach with 25 and 81 sets of multiphysics simulation data, respectively.} Remarkably, the optimization of the overall surrogate model, when trained with 25 sets of samples, converged in only 8 iterations, achieving the optimal solution that met all the tuning states. The optimal design parameter values are represented as \(\bm{x}^* = [3.52961 \; 4.31686\; 3.15937\; 2.91082\; 439.578\; 312.982\; 90.0446\; -80.076\; -309.37\; -598]^T\). The optimization process, when trained with 81 sample sets, converged in 6 iterations to achieve the optimal solutions, denoted as \(\bm{x}^* = [3.56204 \; 4.319\; 3.15958\; 2.91397\; 439.578\; 319.541\; 77.8254\; -80.076\; -368.258\; -598]^T\). Table~\ref{Ex2_Comparisons} presents a comparative analysis between the technique proposed in this article and existing methods for designing the tunable evanescent mode cavity filter {\cite{Tunable_Filter6}}. When an equal number of samples (specifically, 25 training sample sets) are employed, our proposed method yields a more precise overall surrogate model. {The proposed method reduces the total consumption time by 21.6\% compared to the method using ANN model trained by 25 sets of multiphysis samples, and 49.7\% compared to the method using ANN model trained by 81 sets of multiphysis samples.} Furthermore, our optimization approach reaches the optimal solution in fewer iterations than existing techniques do. Compared with other methods that utilize 81 training sample sets, our proposed method can achieve a similarly accurate overall surrogate model and identify the optimal solution more rapidly when only 25 training sample sets are used. The objective function values throughout all iterations for both the proposed and existing techniques are depicted in Fig. \ref{Ex2_Values_of_objective_function}. Our proposed optimization converges within 5 iterations, attaining the optimal solution across all the tuning states with fewer iterations than the two existing techniques do. Consequently, our proposed method demonstrates greater efficiency than current techniques do.

\section{Conclusion}

This article introduces a novel space-mapping-based surrogate-assisted multiphysics optimization technique for the design of microwave tunable filters with multiple tuning states. {This method involves conducting a multiphysics analysis on a microwave tunable filter, accounting for the piezoelectric effects generated by input voltages. The overall surrogate model is composed of multiple subsurrogate models, each of which consists of one shared coarse model and two distinct mapping neural networks. The EM single-physics response of the shared coarse model, established via an ANN, provides a wealth of knowledge and a good approximation for the multiphysics response of the fine model. The mapping neural network translates the EM filed (single-physics filed) into the multiphysics filed. Two mapping functions are employed to minimize the misalignment caused by the piezoelectric effect between filed. }The proposed optimization technique constrains the values of nontunable design parameters by all tuning states, whereas the values of tunable design parameters are constrained only by the corresponding tuning state. All the tuning states are considered and optimized simultaneously, and all the subsurrogate models share a coarse model, thereby conserving computer resources. The established overall model can also be used for high-level EM-centered multiphysics filed design and optimization. The tunable filter design methodology proposed herein is not constrained by the tuning range. Future studies could apply this technique to address microwave structures with an increased number of tuning states. {Another possible future direction would be applying the proposed techniques to thermal-electrical coupled multiphysics examples.} Possible future directions are to explore extrapolation techniques and also consider measurement data for training and testing of the multiphysics parametric model.
\ifCLASSOPTIONcaptionsoff
\newpage
\fi

\begin{IEEEbiography}[{\includegraphics[width=1in,height=1.25in,clip,keepaspectratio]{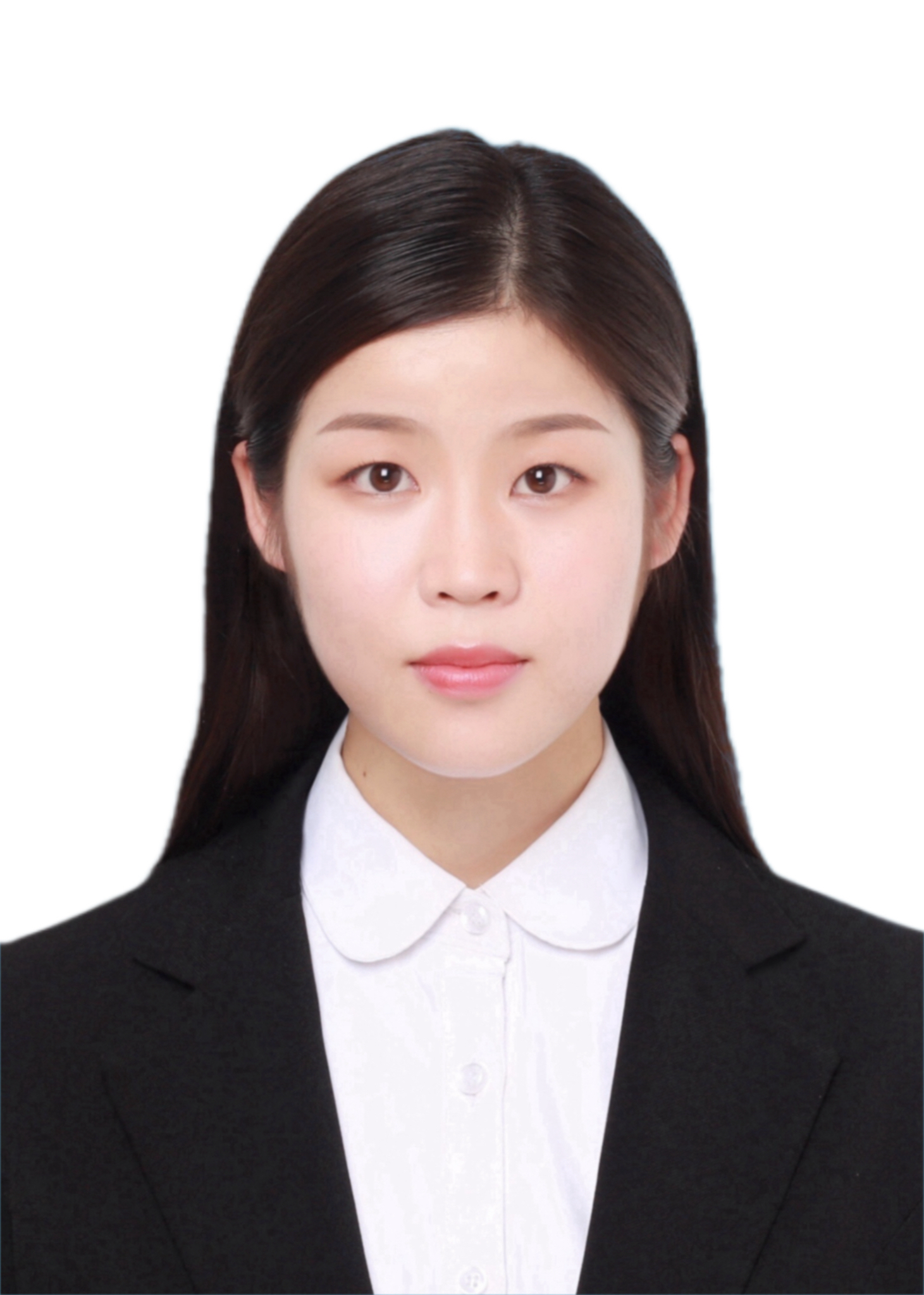}}]{Haitian Hu}(Student Member, IEEE) 
	received the B.Eng. degree from Liaocheng University, Shandong, China, in 2018. She is currently pursuing the Ph.D. degree at School of Electronic Engineering, Beijing University of Posts and Telecommunications, Beijing, China.
	
	Her research interests include neural network-based parametric modeling, electromagnetic-centric multiphysics modeling and optimization, space mapping algorithms, surrogate modeling and surrogate assisted optimization, and multiphysics modeling and optimization.
\end{IEEEbiography}

\begin{IEEEbiography}[{\includegraphics[width=1in,height=1.25in,clip,keepaspectratio]{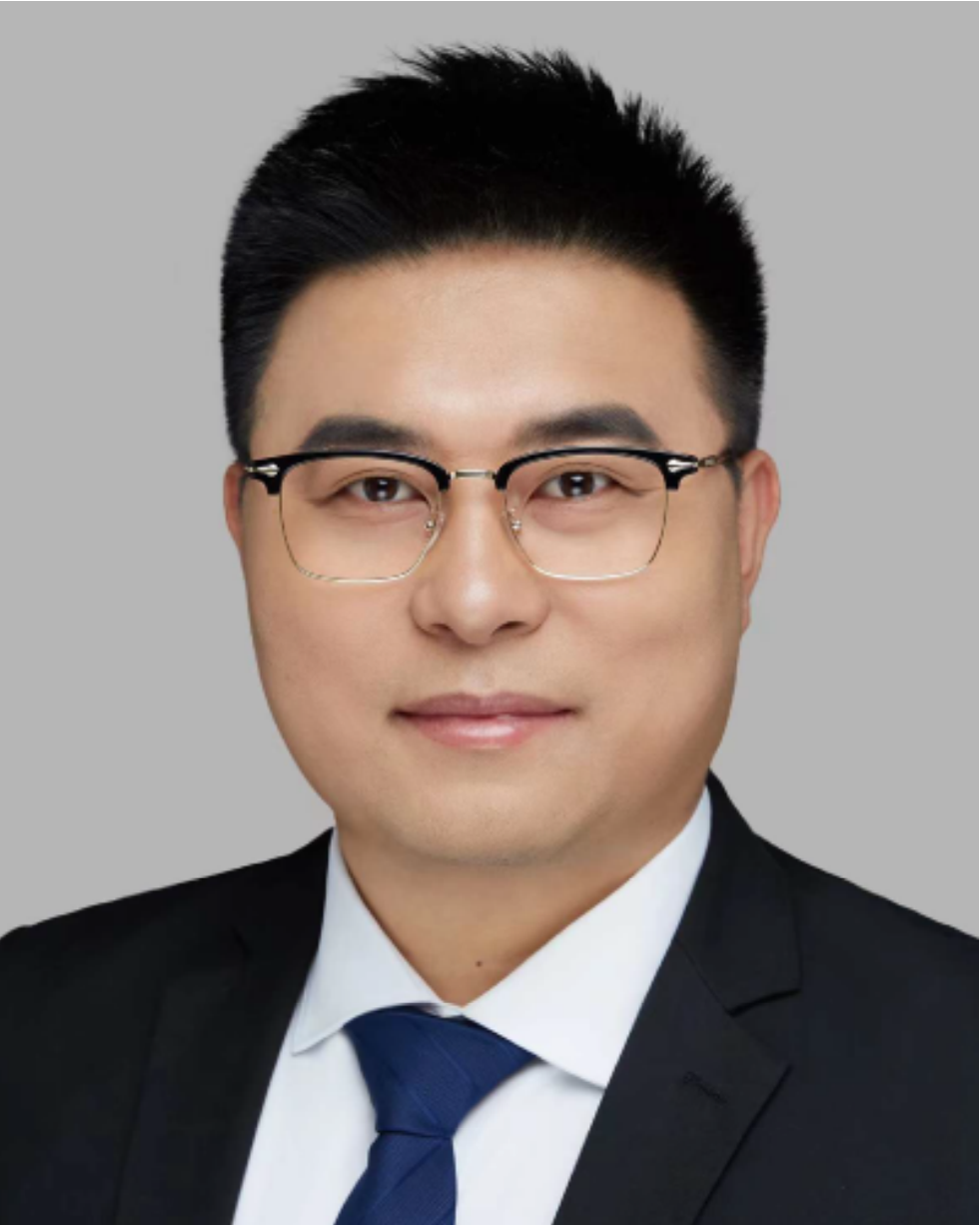}}]{Wei Zhang}
	(Member, IEEE) received the B.Eng. degree from Shandong University, Shandong, China, in 2013. He received the Ph.D. degree from the School of Microelectronics, Tianjin University, and the Department of Electronics, Carleton University, Ottawa, ON, Canada, in 2020. From 2020 to 2021, he was a Postdoctoral Fellow in the Department of Electronics at Carleton University, Ottawa, ON, Canada. 

    He is currently an Associate Researcher with the School of Electronic Engineering at Beijing University of Posts and Telecommunications, Beijing, China. His research interests include neural-network-based methods for microwave device modeling, electromagnetic centric multiphysics modeling and optimization, space mapping algorithm and surrogate model optimization, and computational electromagnetics.
\end{IEEEbiography}

\begin{IEEEbiography}[{\includegraphics[width=1in,height=1.25in,clip,keepaspectratio]{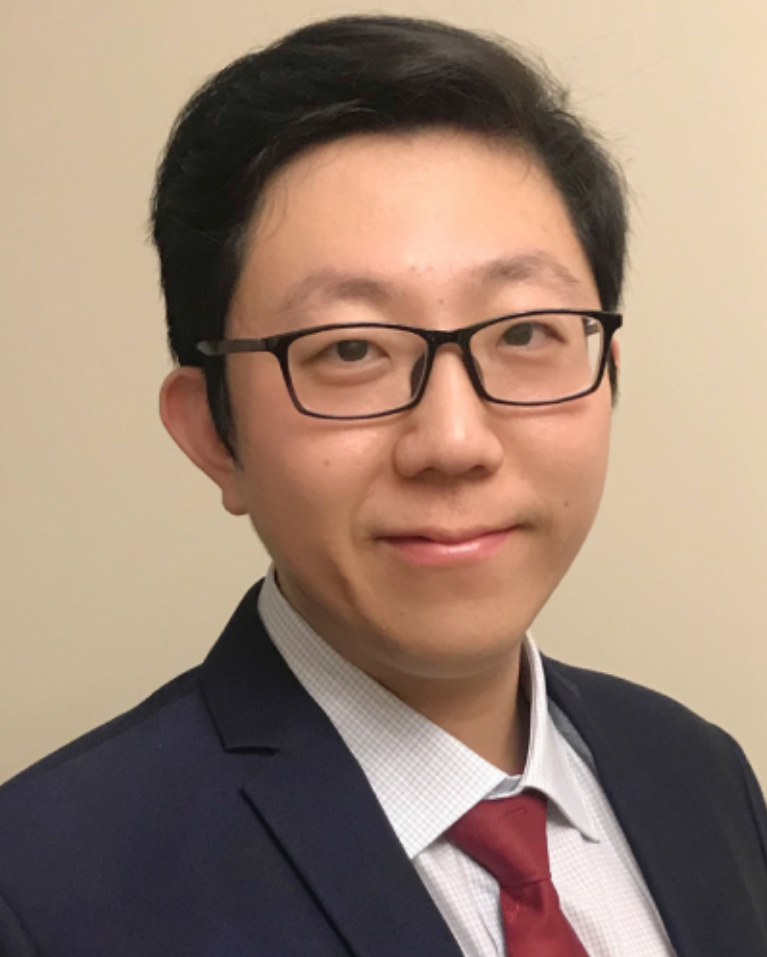}}]{Feng Feng}
	(Senior Member, IEEE)  received the B.Eng. degree from Tianjin University, Tianjin, China, in 2012, and the Ph.D. degree from the School of Microelectronics, Tianjin University, and the Department of Electronics, Carleton University, Ottawa, ON, Canada, in 2017. 
	
	From 2017 to 2020, he was a Post-Doctoral Fellow with the Department of Electronics, Carleton
	University, Ottawa, ON, Canada. He is currently
	an Associate Professor with the School of Microelectronics, Tianjin University. His research interests
	include electromagnetic parametric modeling and design optimization algorithms, deep neural network modeling methods, space mapping algorithm
	and surrogate model optimization, finite-element method in electromagnetic
	simulation and optimization, and multiphysics modeling and optimization.
\end{IEEEbiography}

\begin{IEEEbiography}[{\includegraphics[width=1in,height=1.25in,clip,keepaspectratio]{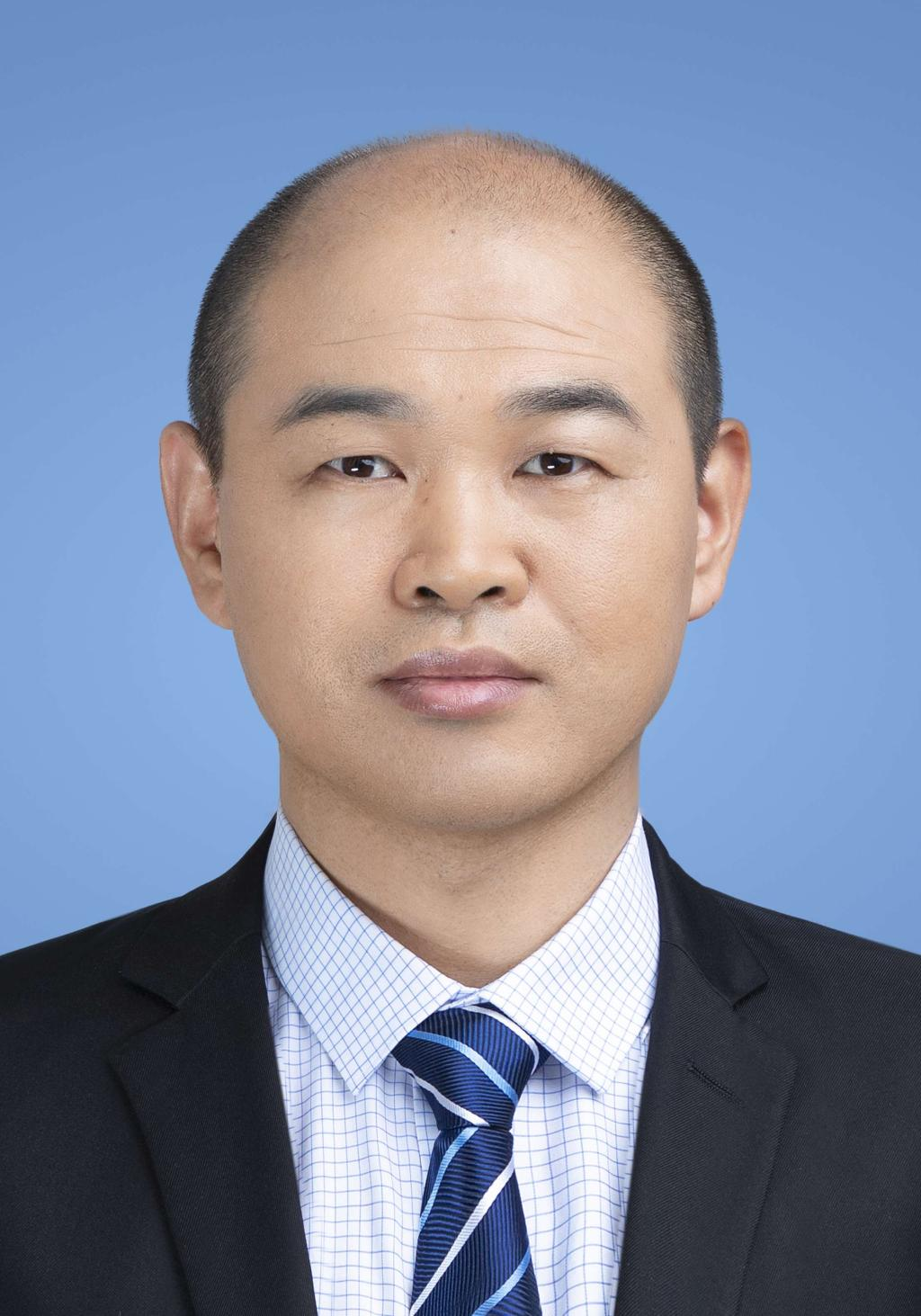}}]{Zhiguo Zhang}
	received the B.Eng. degree from Shandong University, Shandong, China, in 2002. In 2007, He received the Ph.D. degree from the School of Electronic Engineering, Beijing University of Posts and Telecommunications, in 2007.
	
    He is currently a Professor with the School of Electronic Engineering at Beijing University of Posts and Telecommunications, Beijing, China. His research interests include finite-element analysis in electromagnetics, neural network-based methods for nonlinear microwave device and circuit modeling and their applications in computer-aided design for electronic circuits.
\end{IEEEbiography}

\begin{IEEEbiography}[{\includegraphics[width=1in,height=1.25in,clip,keepaspectratio]{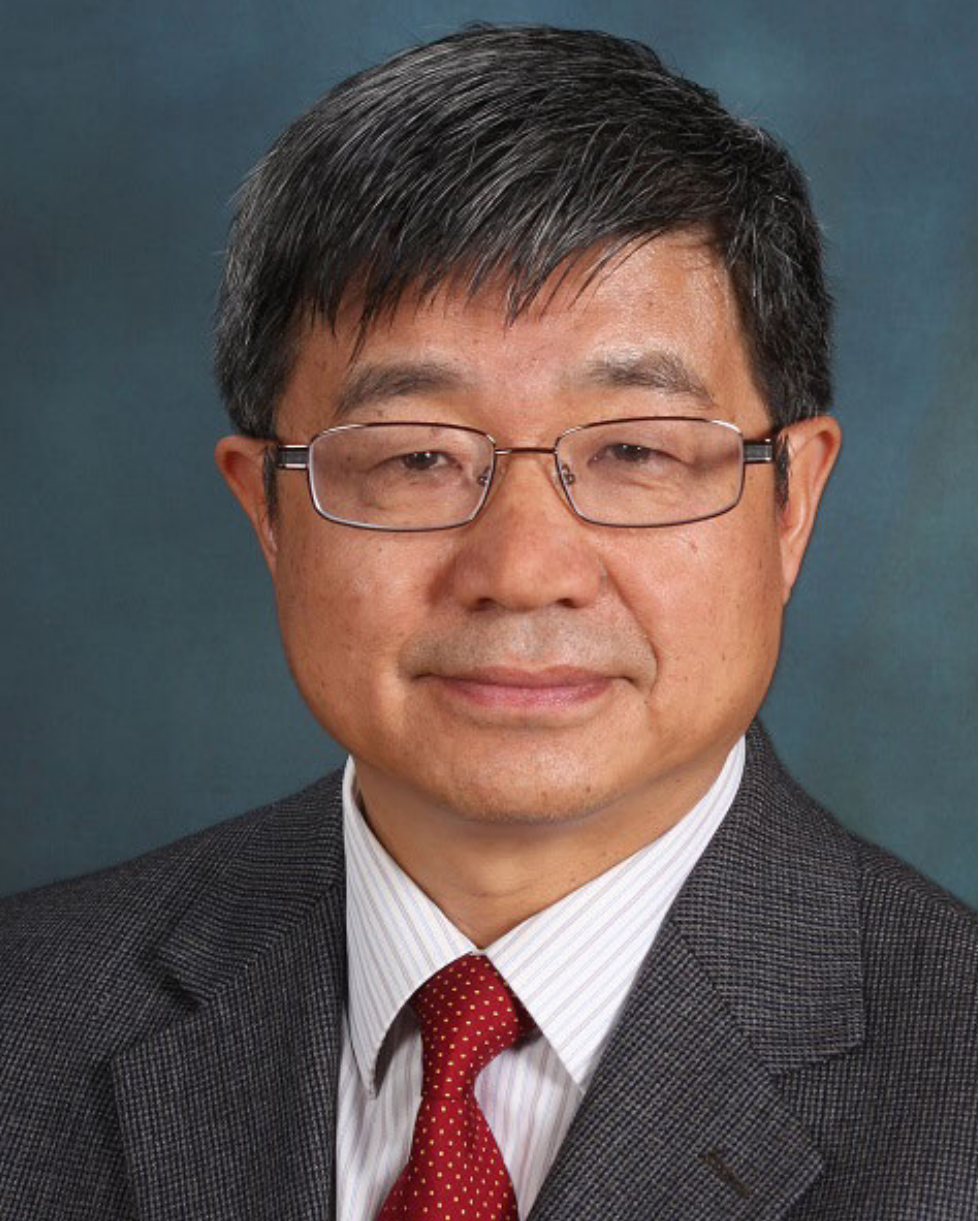}}]{Qi-Jun Zhang}
	(Fellow, IEEE) received the B.Eng. degree from the Nanjing University of Science and Technology, Nanjing, China, in 1982, and the Ph.D. degree in electrical engineering from McMaster University, Hamilton, ON, Canada, in 1987.
	
	He was a Research Engineer with Optimization Systems Associates Inc., Dundas, ON, Canada, from 1988 to 1990, developing advanced optimization software for microwave modeling and design. In 1990, he joined the Department of Electronics, Carleton University, Ottawa, ON, Canada, where he is currently a Chancellor’s Professor. He is an author of the book \emph{Neural Networks for RF and Microwave Design} (Boston, MA, USA: Artech House, 2000) and a co-editor of the book \emph{Modeling and Simulation of High-Speed VLSI Interconnects} (Boston, MA, USA: Kluwer, 1994) and \emph{Simulation-Driven Design Optimization and Modeling for Microwave Engineering} (London, U.K.: Imperial College Press, 2013). His research interests include modeling, optimization, and machine learning for high-speed/high-frequency electronic design. He has more than 360 publications in the area.
	
	Dr. Zhang is a Fellow of the Canadian Academy of Engineering and the Engineering Institute of Canada. He was twice a Guest Editor of the Special Issues on Applications of ANN for RF/Microwave Design for the \emph{International Journal of RF/Microwave Computer-Aided Engineering} in 1999 and 2002, and a Guest Co-Editor of the Special Issue on Machine Learning in Microwave Engineering for the \emph{IEEE Microwave Magazine} in 2021. He is an Associate Editor of the \textsc{IEEE Transactions on Microwave Theory and Techniques} and a Topic Editor of the \textsc{IEEE Journal of Microwaves}. He is also the Co-Chair of the Working Group on AI and Machine Learning Based Technologies for Microwaves in the Future Directions Committee of the IEEE Microwave Theory and Techniques (MTT) Society.
\end{IEEEbiography}

\end{document}